\documentclass[a4paper,fleqn,usenatbib]{mnras}

\usepackage[T1]{fontenc}
\usepackage{ae,aecompl}

\usepackage{graphicx}	
\usepackage{amsmath}	
\usepackage{amssymb}	
\usepackage{color}

\title[Analysis of a triply eclipsing system]{Photodynamical analysis of the triply eclipsing hierarchical triple system EPIC\,249432662}

\author[T. Borkovits et al.]{
T. Borkovits$^{1,2}$, 
S.~Rappaport$^3$,   
T. Kaye$^4$, 
H. Isaacson$^5$, 
A. Vanderburg$^{6,7,8}$, 
\newauthor
A.W. Howard$^9$, 
M.H. Kristiansen$^{10,11}$, 
M.R. Omohundro$^{12}$,
H.M. Schwengeler$^{12}$,
\newauthor
I.A. Terentev$^{12}$,
A. Shporer$^{13}$, 
H. Relles$^{13}$, 
S. Villanueva Jr.$^{13,14}$, 
T.G. Tan$^{15}$, 
\newauthor
K.D. Col\'on$^{16}$
J. Blex$^{17}$,
M. Haas$^{17}$,
W. Cochran$^{7}$,
and M. Endl$^{7}$
\\
$^1$ Baja Astronomical Observatory of Szeged University, H-6500 Baja, Szegedi \'{u}t, Kt. 766, Hungary; borko@electra.bajaobs.hu \\
$^2$ Konkoly Observatory, Research Centre for Astronomy and Earth Sciences, Hungarian Academy of Sciences, \\
Konkoly Thege Miklós \'ut 15-17, H-1121 Budapest, Hungary \\
$^3$ Department of Physics, and Kavli Institute for Astrophysics and Space Research, M.I.T., Cambridge, MA 02139, USA; sar@mit.edu \\
$^4$  Raemor Vista Observatory, 7023 E. Alhambra Dr., Sierra Vista, AZ 85650, USA\\
$^5$ Department of Astronomy, University of California at Berkeley, Berkeley, CA 94720-3411, USA\\
$^6$ Harvard-Smithsonian Center for Astrophysics, 60 Garden Street, Cambridge, MA 02138 USA; avanderburg@cfa.harvard.edu \\ 
$^7$ Department of Astronomy, The University of Texas at Austin, 2515 Speedway, Stop C1400, Austin, TX 78712 \\
$^8$ NASA Sagan Fellow \\
$^9$ Astronomy Department, California Institute of Technology, MC 249-17, 1200 E. California Blvd, Pasadena, CA 91125, USA \\
$^{10}$ DTU Space, National Space Institute, Technical University of Denmark, Elektrovej 327, DK-2800 Lyngby, Denmark \\
$^{11}$ Brorfelde Observatory, Observator Gyldenkernes Vej 7, DK-4340 T\o ll\o se, Denmark \\
$^{12}$ Citizen Scientist\\
$^{13}$ Kavli Institute for Astrophysics and Space Research, M.I.T., Cambridge, MA 02139, USA \\
$^{14}$ Department of Astronomy, The Ohio State University, Columbus, OH 43210, USA\\
$^{15}$ Perth Exoplanet Survey Telescope, Perth, Western Australia 6010 \\
$^{16}$ NASA Goddard Space Flight Center, Exoplanets and Stellar Astrophyscs Laboratory (Code 667), Greenbelt, MD 20771, USA\\
$^{17}$ Astronomisches Institut, Ruhr Universit\"at, 44780 Bochum, Germany\\
}

\date{Accepted XXX. Received YYY; in original form 2018 September 3}

\pubyear{2018}

\begin{document}
\label{firstpage}
\pagerange{\pageref{firstpage}--\pageref{lastpage}}
\maketitle

\begin{abstract}
Using Campaign 15 data from the {\em K2} mission, we have discovered a triply-eclipsing triple star system: EPIC\,249432662.  The inner eclipsing binary system has a period of 8.23 days, with shallow $\sim$3\% eclipses.  During the entire 80-day campaign, there is also a single eclipse event of a third-body in the system that reaches a depth of nearly 50\% and has a total duration of 1.7 days, longer than for any previously known third-body eclipse involving unevolved stars.  The binary eclipses exhibit clear eclipse timing variations.  A combination of photodynamical modeling of the lightcurve, as well as seven follow-up radial velocity measurements, has led to a prediction of the subsequent eclipses of the third star with a period of 188 days.  A campaign of follow-up ground-based photometry was able to capture the subsequent pair of third-body events as well as two further 8-day eclipses.  A combined photo-spectro-dynamical analysis then leads to the determination of many of the system parameters.   The 8-day binary consists of a pair of M stars, while most of the system light is from a K star around which the pair of M stars orbits.  
\end{abstract}




\section{Introduction}

Hierarchical triple stellar systems and/or subsystems form a small but very important subgroup of the zoo of multiple stellar systems.
Their significance in the formation of the closest main-sequence binary systems \citep[see, e. g.][and further references therein]{eggletonkiseleva-eggleton01,fabryckytremaine07,naozfabrycky14,maxwellkratter18} is widely acknowledged. Hierarchical triples have also been hypothetised to play a significant role in the formation of some kinds of peculiar objects, such as blue stragglers \citep{peretsfabrycky09}, low-mass X-ray binaries \citep{shappeethompson13}, and some peculiar binary pulsars \citep[see, e. g.][]{portegieszwartetal11}. They might even have a role in driving two white dwarfs to merger in that scenario for type Ia supernova explosions \citep{maozetal14}.

\begin{figure}
\includegraphics[width=0.97 \columnwidth]{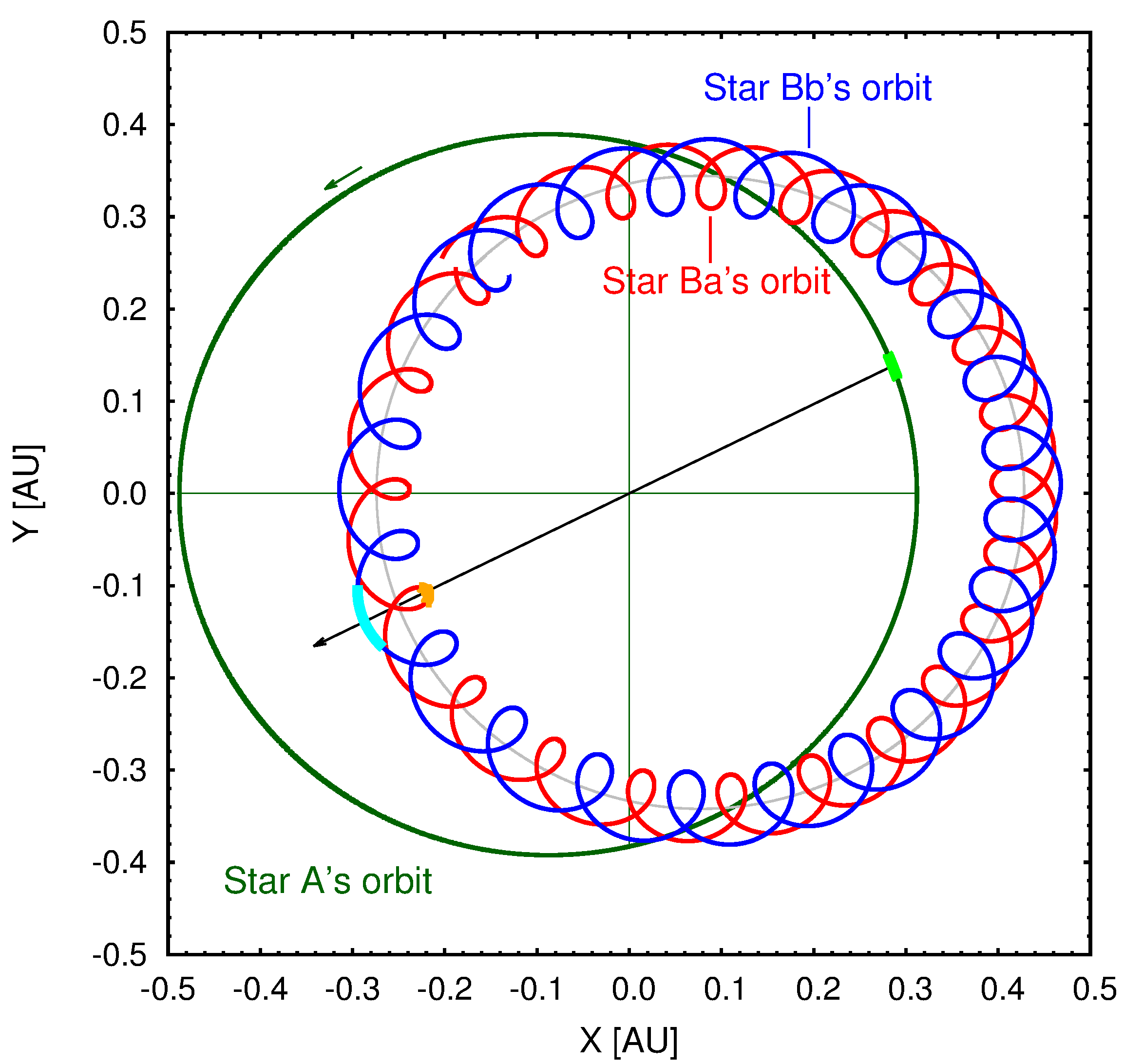}
\caption{Pole-on view of the hierarchical triple star system EPIC\,249432662. The red and blue curves represent the motions of the Ba and Bb stars of the `inner' 8-day binary orbit in their 188-day `outer' orbit about the center of mass (CM) of the triple system (located at $X=Y=0$).  The thin grey curve marks the locus of CM points for the 8-day binary.  The green curve is the 188-day `outer' orbit of star A, the third star which comprises the system. The thin green lines denote the major and minor axes of the orbit of star A, while the thin green arrow indicates the direction of motion along the orbits.  Thicker sections of the orbits represent the arcs on which the three stars were moving during the `great eclipse', observed with the {\em Kepler} spacecraft around BJD\,2\,458\,018. The black arrow which connects these arcs is directed toward the observer.}
\label{fig:CHTscheme} 
\end{figure}

In the case of {\em hierarchical} triples, one of the three mutual distances among the three components of the system remains substantially smaller than the other two distances for the entire lifetime of the triple. Therefore, the dynamics of the triple can be well approximated with the (slightly perturbed) Keplerian motions of two `binaries': an inner or close binary, formed by the two components having the smallest separation, and an outer or wide `binary', consisting of the centre of mass of the inner pair and the more distant third object (see Fig.\,\ref{fig:CHTscheme} for a schematic diagram)\footnote{In this paper, we use the following notations. The orbital elements of the inner and outer orbits are subscripted by numbers `1' and `2', respectively. Regarding the three stars, we label them as $A$, $Ba$, and $Bb$, where $A$ denotes the brightest and most massive component, i.e., the distant, third star, while $Ba$ and $Bb$ refer to the primary and the secondary components of the close, inner 8-day binary. When we refer to physical quantities of individual stars, we use these subscripts. In such a way, for example, $m_\mathrm{A}$ or $m_\mathrm{Ba}$ stands for the masses of components $A$ or $Ba$, respectively, but $m_\mathrm{B}$ denotes the total mass of the inner binary, ($m_\mathrm{Ba}+m_\mathrm{Bb}$), while $m_\mathrm{AB}$ refers to the total mass of the entire triple system.}. For most of the known hierarchical triple star systems, with large period ratios for the `outer binary' to `inner binary', departures from pure Keplerian motion are expected to become significant only on timescales of decades, centuries, or even millenia, i.e., over much longer intervals than the length of the available observational data trains.  However, there is an important subgroup of the hierarchical triple systems, the so-called `compact hierarchical triples' (CHT) which have smaller ratios of `outer' to `inner' periods and, occasionally, also smaller characteristic orbital dimensions.  These may show much shorter timescale and well-observable dynamical (or other kinds of) interactions so that they allow us to promptly determine many of the important dynamical and astrophysical parameters of these systems.

For example, a careful analysis based (partly) on the dynamically perturbed pulsar timing data of the millisecond pulsar PSR\,J0337+1715 orbiting in a peculiar CHT consisting of two white dwarfs in addition to the pulsar component, has led to the accurate determination of the masses of all three objects, as well as the spatial configuration of the triple \citep{ransometal14}. Similarly, as was shown by \citet{borkovitsetal11,borkovitsetal15}, if the close pair of a CHT happens to be an eclipsing binary (EB) the dynamical perturbations of the third companion on the orbital motion of the EB manifest themselves in intensive and quasi-cyclic eclipse timing variations (ETV) on the time-scale of the orbital period of the outer component. The analysis of this effect makes it possible to determine not only the complete spatial configuration of the triple system, but also the masses of the three objects. 

As an application of the latter ETV analysis method, \citet{borkovitsetal16} investigated 62 such CHTs in the original field of the {\em Kepler} space telescope \citep{boruckietal10} where the inner binary was an EB, and the dynamical interactions were significant. They were able to determine many of the system parameters, including the mutual inclination of the planes of the inner and outer orbits, which is a key parameter from the point of view of the different triple star formation theories \citep[see, e. g.][]{tokovinin17}. In the same paper the authors identified an additional 160 CHTs through the traditional light-travel time effect \citep[see, e.g.][]{irwin59} and, in total, they found $\sim$$104$ CHTs with outer orbital period $P_2\lesssim1000$\,d; this provides by far the most populated sample of hierarchical triple stars at the lower end of their outer period distribution. They pointed out a significant dearth of ternaries with $P_2\lesssim200$\,d, and concluded that this fact cannot be explained with observational selection effects. This latter result is in accord with the previous findings of \citet{tokovininetal06} on the unexpected rarity of (the mostly spectroscopically discovered) triple systems in the period regime $P_2<1000$\,d, whose shortage is more explicit amongst those CHTs which contain exclusively solar mass and/or less massive components \citep{tokovinin14}. Therefore, investigations of such systems are especially important.

A very narrow subgroup of CHTs that offers further extraordinary possibilities for accurate system parameter determinations are those triples, which exhibit outer eclipses. These systems have a fortuitous orientation of the triple system relative to the observer whereby, occasionally, the distant third component eclipses one or both stars of the inner close binary or, vice versa, it is eclipsed by them.  Such phenomena had never been seen before the advent of the {\em Kepler} era. The {\em Kepler} space telescope's 4-year-long, quasi-continuous observations, made with unprecedented photometric precision have, however, led to the discovery of at least 11 CHTs exhibiting outer eclipses, and a similar number of circumbinary transiting extrasolar planets. The latter group, though dynamically similar, are not considered in the following list of CHTs. These are KIC\,05897826 \citep[=KOI-126;][]{carteretal11}, KIC\,05952403 \citep[=HD\,181068;][]{derekasetal11}, KICs\,06543674, 07289157 \citep{slawsonetal11}, KIC\,02856960 \citep{armstrongetal12,marshetal14}, KIC\,02835289 \citep{conroyetal14}, KICs\,05255552, 06964043, 07668648 \citep{borkovitsetal15}, KIC\,09007918 \citep{borkovitsetal16}, and KIC\,0415061 (=HD\,181469) the latter of which is possibly at least a quintuple system \citep{TheThing,helminiaketal17}. For ten of these eleven CHTs, besides the outer eclipses, the inner binary also shows regular eclipses; hence, we call these CHTs `triply eclipsing systems'. (The only exception is KIC\,02835289, where the inner binary is an ellipsodial light variable.) Though the much shorter duration of the {\em K2} observations is less favourable in regard to the discovery of systems with outer eclipses, another triply eclipsing CHT, HD\,144548 was also identified in the C2 field of the extended {\em K2} mission \citep{alonsoetal15}. Furthermore, recently, \citet{hajduetal17} reported the discovery of two additional triply eclipsing CHTs, namely CoRoTs\,104079133 and 221664856 amongst the EBs observed by the CoRoT space telescope \citep{auvergneetal09}.

Precise modeling of the brightness variations of these CHTs, especially during each outer eclipse, is a great challenge but, on the other hand, it offers huge benefits. This is so because the lightcurve is extremely sensitive to the complete configuration of the triple.  As a consequence, even in the case where the outer orbit is wide enough to safely allow for the elimination of any dynamical perturbations in the analysis of accurate ETV and/or radial velocity (RV) curves, the same cannot be done for the lightcurve solution. This is true because dynamically induced departures in the positions and velocities of the three bodies relative to a purely Keplerian motion, even if they are very small, will strongly affect all the characteristics of the forthcoming outer eclipses. Therefore, the accurate modeling of such systems, in most cases, requires a photodynamical approach, including the complete numerical integration of the motion of the three bodies, together with the simultaneous analysis of the lightcurve(s) (and, if it is available, the RV curve[s], as well), as was carried out by, e.g.,\,\cite{carteretal11}, for KOI-126 and \citet{orosz15}, for KIC 07668648.\footnote{In addition to KIC\,07668648, J.~Orosz has also successfully applied the same combined photodynamical approach to other CHTs amongst the {\em Kepler}-discovered EBs (e.\,g. to KIC\,10319590) that did not exhibit outer eclipses, but the rapid eclipse depth variations and the features of the ETV and RV curves have allowed him to infer accurate system parameters.}

In this paper we report the discovery, and present the photodynamical analysis, of the new triply eclipsing CHT EPIC\,249432662 (=2MASS\,J15334364-2236479, UCAC4\,337-074729). The system was observed during Campaign 15 of the {\em K2} mission. Besides the regular, $\sim$$3$\%-deep eclipses belonging to an $8.23$-day period, slightly eccentric EB, the {\em Kepler} spacecraft has observed an additional, 1.7-day-long, irregularly shaped fading event with an amplitude of almost 50\% which we assumed to be an outer eclipse due to the presence of a third, distant, gravitationally bound stellar component. Ground-based spectroscopic and photometric follow-up observations have confirmed our assumption and made it possible to carry out a complete, joint photo-spectro-dynamical analysis of this 188.3-day-outer-period, triply eclipsing CHT, including the simultaneous joint analysis of {\em K2} and ground-based lightcurves, the ETV curve (derived from the photometric observations), and the ground-based RV curve, all accompanied by the numerical integration of the motion of the three bodies. We derive many of the parameters for this system. The paper is organized as follows. In Sect.~\ref{sec:K2} we describe the 80-day {\em K2} observation of EPIC\,249432662. Existing archival data on the target star are summarized in Sect.~\ref{sec:archive}. Our two ground-based photometric follow up campaigns, which led to the successful detection of further outer eclipses, are discussed in Sect.~\ref{sec:photometry}, while the ETV data determined from both the {\em K2} and ground-based photometry is briefly described in Sect.\,\ref{sec:ETV}. In Sect.\,\ref{sec:RV} we present our spectroscopic RV follow-up observations.  We then use our improved photodynamical software package to model and evaluate the satellite and ground-based lightcurves, ETV curves, and RV results simultaneously (see Sect.~\ref{sec:model}). In Sect.~\ref{sec:discussion} we discuss some of the astrophysical and orbital/dynamical implications of our solutions.  We summarize our findings and draw some conclusions in Sect.~\ref{sec:conclusion}.  Finally, we discuss some practical details of our photodynamical code in Appendices \ref{app:numint} and \ref{app:ETV}.

\section{{\em K2} Observations}
\label{sec:K2}

Campaign 15 (C15) of the {\em K2} mission \citep{howell14} was directed toward the constellation Scorpius between 2017 August 23 - 2017 November 20 for approximately 87 uninterrupted days. At the end of November, the {\em K2} Guest Observer (GO) Office made the C15 raw cadence pixel files (RCPF) publicly available on the Barbara A.~Mikulski Archive for Space Telescopes (MAST)\footnote{\url{http://archive.stsci.edu/k2/data_search/search.php}}. We utilized the RCPF in conjunction with the Kadenza software package\footnote{\url{https://github.com/KeplerGO/kadenza}}  \citep{kadenza}, in a manner similar to recent {\em K2}-discoveries \citep[see, e.g.,][]{christiansen, yu,david}, combined with custom software in order to generate minimally corrected light curves. Due to limited processing power, some of us (MO, IT, HMS, MHK) generated a series of short-baseline C15 preview light curves, and carried out manual surveys with the {\tt LcTools} software \citep{Kipping}. The first C15 preview search identified EPIC\,249432662 (Proposed by: GO15083, Coughlin)\footnote{\url{https://keplerscience.arc.nasa.gov/data/k2-programs/GO15083.txt}}  as a likely 8-d eclipsing binary, while our second extended preview light curve identified an additional single, deep, compound eclipsing event of long duration centered at BJD = 2458018.  

\begin{figure}
\includegraphics[width=0.97 \columnwidth]{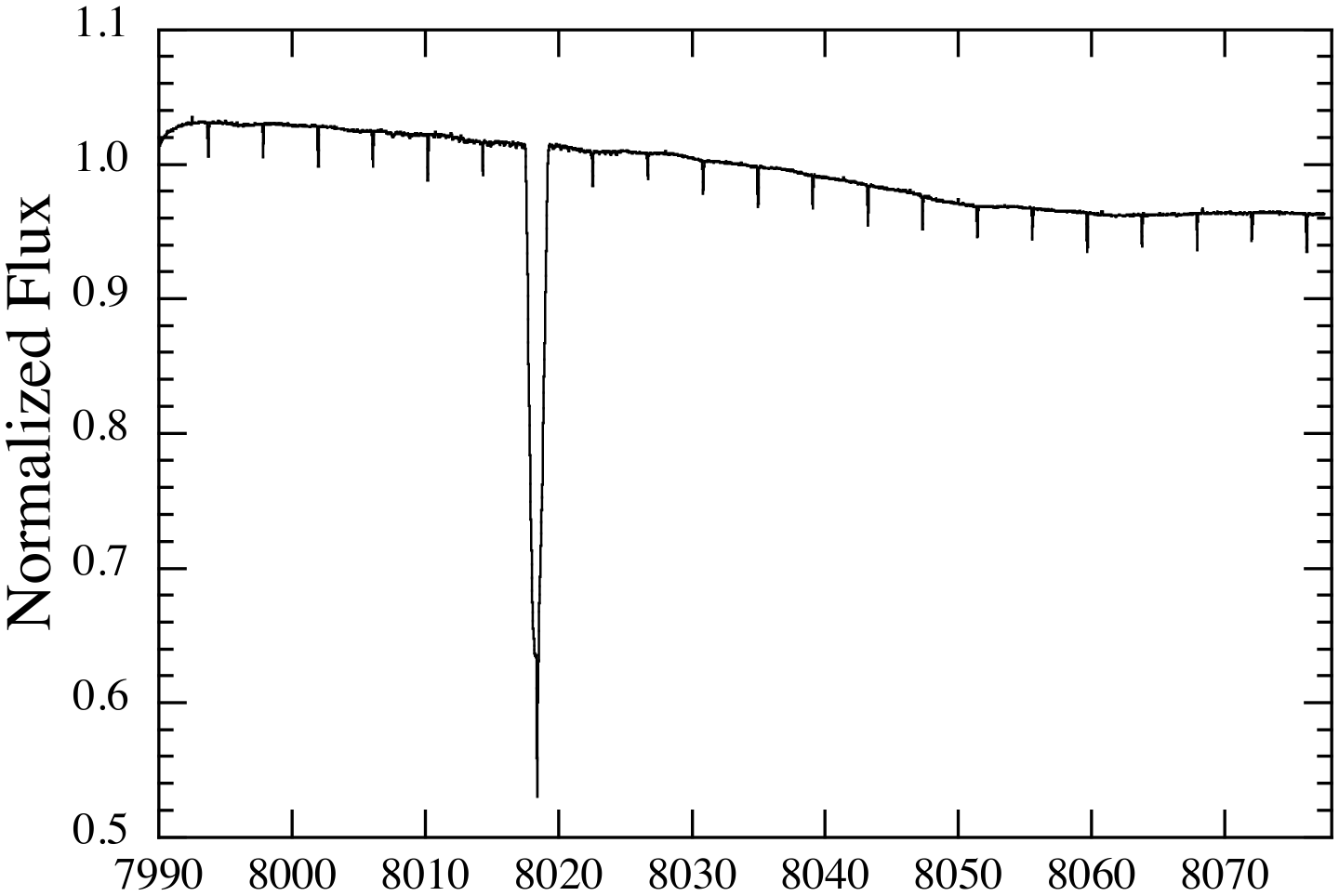} \includegraphics[width=1.00 \columnwidth]{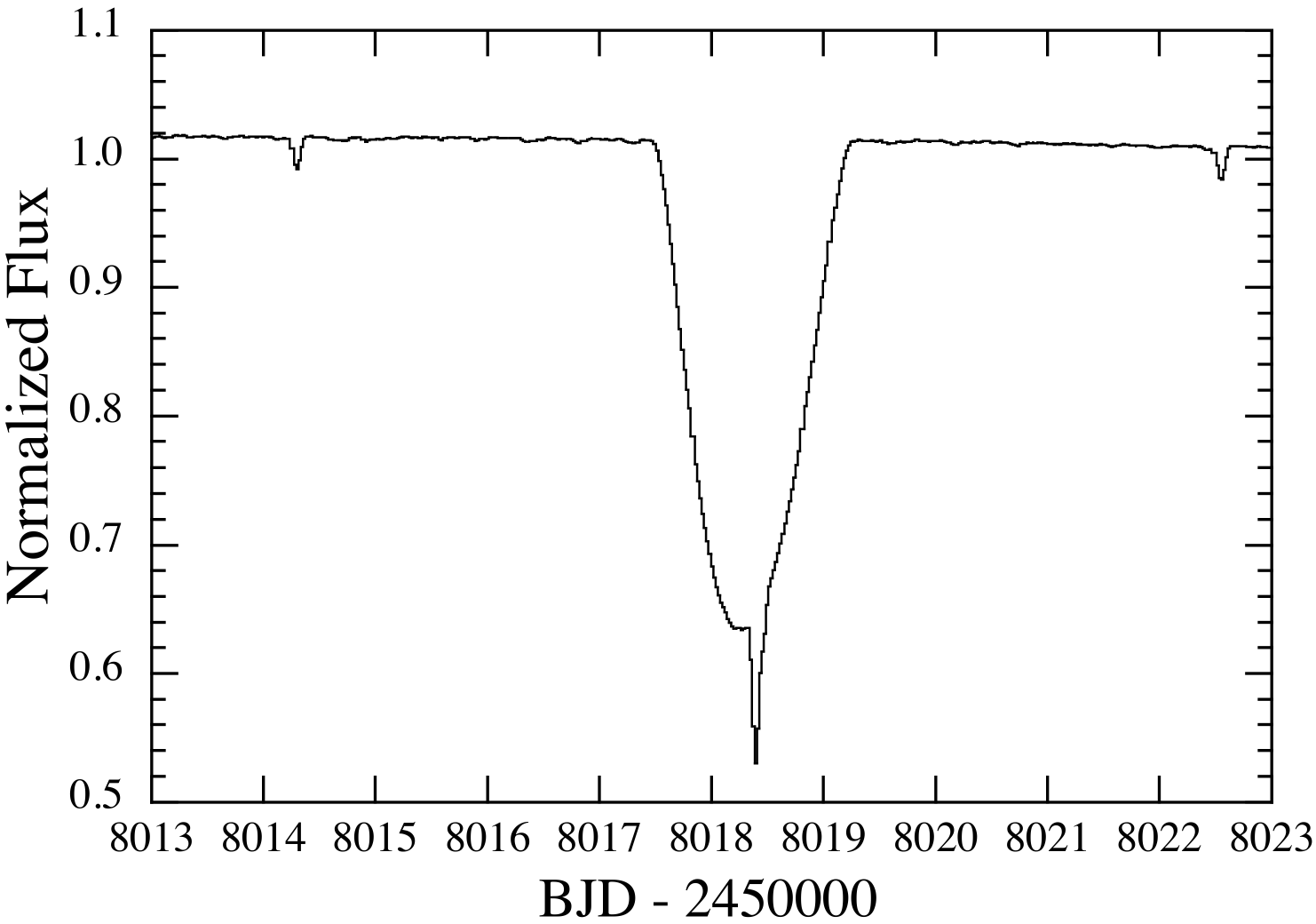}
\caption{{\em K2} lightcurve of EPIC\,249432662 from Campaign 15. {\em Top panel}: full 80-day lightcurve; {\em Bottom panel:} 10-day zoom-in around the long, deep, and structured eclipse of the third star by the inner binary.}
\label{fig:lightcurve} 
\end{figure}

Once the calibrated Ames data set was released, we downloaded all available {\em K2} extracted light curves common to Campaign 15 from the MAST. We utilized the pipelined data set of \citet{vanderburg2014} to construct the lightcurve, which is presented in Fig.~\ref{fig:lightcurve}; the top panel being the data for all 80 days of the {\em K2} observation, whereas the bottom panel shows a 10-day zoom in around the large eclipse which we dub the ``great eclipse'' (or `GE').  The two nearly equal-depth eclipses from the 8-day binary are also clearly evident.  When looked at in the expanded view, the ``great eclipse'' is seen to be comprised of a deep and slightly asymmetric portion (here called ``GE1''), and a sharp (i.e., short-duration) extra dip (called ``GE2'') near the minimum of the event.

Note, however, that in our photodynamical analysis (see Sect.\,\ref{sec:model}) we used a ``flattened'' version of the lightcurve that differs from the one shown in Fig.~\ref{fig:lightcurve} by having the long-term trend and low-frequency variability removed.  The procedure is to iteratively fit a basis spline (B-spline) with breakpoints every 1.5 days to the lightcurve with the 3-$\sigma$ outliers (including, of course, the eclipses) removed from the fit.  This process is repeated until convergence is achieved (see \citealt{vanderburg2014}). The eclipses, including the great eclipse, are then added back to the spline fit.

We have examined both the {\em K2} pixel-level data and the PANStarrs image \citep{chambers16} of the field to check on `third-light' contamination to the {\em K2} light curve from neighbor stars. We utilized the pixel reference function of {\em Kepler} (on the module with the Great Eclipser; \citealt{bryson10}), which gives the contribution from a given star as a function of distance from the maximum in flux.  The nearest star that could add contaminating light is 15$''$ away and is only $\sim$20\% the brightness of the target. The contribution from this star, according to the pixel reference function, is then no more than 0.1\% of the target star, and hence negligible.  

Before doing any quantitative analysis, we qualitatively convinced ourselves that GE1 must be due to an eclipse of a third star in the system by one of the binary stars.  This star would just happen to be moving in the same direction on the sky as the third star, and nearly at the same speed, so as to dramatically slow their relative motion and produce an eclipse that lasts for 1.7 days.  By contrast, GE2 must be caused by the other star in the binary which is moving in the {\em opposite} direction on the sky as the third star.  We also concluded early on that the two stars in the binary must be of comparable mass, size, and $T_{\rm eff}$, based on the near equality of the two binary eclipses.  Finally, we realized that there was just a limited set of stellar parameters that would allow for only one star in the binary to be able to block $\sim$38\% of the system light.

In the remainder of the paper we focus on understanding this ``great eclipse'' quantitatively and, in the process, extracting the system parameters.

\section{Archival data}
\label{sec:archive}

The target star image, as a composite of all three stars, has a {\em Kepler} magnitude of 14.93.  The coordinates of the target star and its brightness in other magnitude bands from the blue to the WISE 3 band are summarized in Table \ref{tbl:mags}.  The new Gaia DR2 release puts the target at a distance of $445 \pm 7$ pc.  This distance and the corresponding proper motions from Gaia DR2 are also listed in Table \ref{tbl:mags}.
Note, however, that despite the unprecedented astrometric precision of Gaia, the DR2 parallax, and therefore distance, for the present system should be considered as only a preliminary value, and should not be accepted without some caveats. The reason is that binary star solutions have not yet been incorporated into the DR2 results. In particular, since the orbital period of the outer orbit in our triple is $P_2\sim188$-days, which is almost exactly half of the orbital period of the Gaia satellite, the absence of corrections for the internal motions might be critical for our triple. 

Based on the Gaia photometry and the source distance, the DR2 file on this object lists the star as having a radius of $0.86^{+0.055}_{-0.077} \, R_\odot$.  Of course this analysis is based on the assumption that there is one dominant star present, and the light from the two stars in the 8-day binary do not contribute much to the system light. 

Under these assumptions we know at least that the third star is of K spectral type with a mass of $\sim$$0.7-0.8 \,M_\odot$, and lies quite close to the zero-age main sequence.

\begin{table}
\centering
\caption{Photometric Properties of EPIC\,249432662}
\begin{tabular}{lc}
\hline
\hline
Parameter &
EPIC\,249432662 \\
\hline
Aliases & 2MASS\,J15334364-2236479  \\
& WISE\,J153343.62-223648.1 \\
& UCAC4\,337-074729 \\
& Gaia DR2 6239702584685025280\\
RA (J2000) & 15:33:43.639   \\  
Dec (J2000) &  $-22$:36:48.07  \\  
$K_p$ & 14.93  \\
$B$ & 16.57$^a$ \\  
$V$ & 15.46$^a$ \\
$G$ & 14.92$^b$  \\
$r'$ & 14.95$^a$ \\
$i'$ & 14.45$^a$ \\
J & 12.83$^c$ \\
H & 12.25$^c$ \\
K & 12.09$^c$ \\
W1 & 11.99$^d$ \\
W2 & 11.96$^d$ \\
W3 & 11.71$^d$ \\
Distance (pc) & $445 \pm 7$$^b$  \\   
$\mu_\alpha$ (mas ~${\rm yr}^{-1}$) & $-18.70 \pm 0.07$$^b$  \\ 
$\mu_\delta$ (mas ~${\rm yr}^{-1}$) &  $-10.82 \pm 0.05$$^b$  \\ 
\hline
\label{tbl:mags}
\end{tabular}

{\bf Notes.} (a) UCAC4 \citep{UCAC4}.  (b)  Gaia DR2 \citep{lindegren}. (c) 2MASS archive \citep{Skrutskie}. (d) WISE archive \citep{Cutri}. 
\end{table} 

\section{Ground-Based Follow-up Observations}
\label{sec:photometry}

The history of our ground-based follow-up observations nicely illustrates the role of good fortune in a scientific endeavor. Our preliminary joint photodynamical runs -- including the {\em K2} photometry and the ETV curves derived from it -- clearly demonstrated that the main features of the {\em K2} lightcurve can be well reproduced with a compact, dynamically active, triple stellar system in an almost coplanar configuration. However, due to the strong degeneracies among many of the orbital parameters in terms of the outer eclipsing pattern, we were unable to constrain the outer orbital period and, therefore, to predict the likely time(s) of the forthcoming outer eclipses. This situation changed dramatically after we obtained the fourth RV data point from 2018 March 22. This RV point made it possible to constrain the outer period with only a $\sim$$2-3$-day uncertainty and, therefore, to predict the most probable forthcoming outer eclipse times to within a range of only a few days. The most important consequence was that we then understood that the forthcoming outer eclipses should be occurring within $5-10$ days of that time! Therefore, we had to urgently organize an international observing campaign with several observers from Arizona to Chile.  We were thereby able to perfectly catch the next pair of `great (outer) eclipses' almost exactly after the start of the follow-up ground-based observations.

\vspace{0.2cm}

\noindent
{\em JBO Observations} \\
The observations of the primary third-body eclipses were conducted with the Junk Bond Observatory by author TGK. The telescope is an 80-cm  Ritchie Chretien with an SBIG STL6303E CCD. 60-second unfiltered images were shot sequentially through the events. Darks, flats and data images were reduced using MaximDL software by Bruce Gary.  

The observations were carried out for 5 and 6 hours on the evenings of 2018 March 27 and 29, respectively.  By good fortune, both of the deep primary third-body eclipses were captured photometrically.  

\vspace{0.2cm}

\noindent
{\em RoBoTT Telescope Observations}\\
We have carried out photometry of the target star at the end of 2018 March with the ROBOTT telescope (formerly VYSOS6). The images are taken in two filters: sloan $r'$ and $i'$, during the first night with exposure times of 30 s, and then later at 60\,s.  Typically 9 images are combined with outlier rejection to remove cosmic rays. The images are taken at the same sky position, and the source of interest is at the image center.  While the FoV of the instrument is $2.7 \times 2.7$ degrees, we extracted submaps of $ 45' \times 45'$ FoV centred on the target, and used only stars in this area for the light curve processing.  A description of the data processing and reduction can be found in \citet{haasetal12}.  

\vspace{0.2cm}

\noindent
{\em DEMONEXT Observations}

We obtained additional ground-based photometry for EPIC\,249432662 using the DEMONEXT telescope \citep{Villanueva18} at Winer Observatory in Sonoita, Arizona. DEMONEXT is a 0.5 m PlaneWave CDK20 telescope with a $2048 \times 2048$ pixel FLI Proline CCD3041 camera, a $30.7'' \times 30.7''$ field of view, and a pixel scale of $0.9''$/pixel.

EPIC 249432662 was placed in the DEMONEXT automated queue for continuous monitoring for the night of 2018 March 28. 250 observations were executed while the target was above airmass 2.4. An exposure time of 42 seconds was used and DEMONEXT was defocused to avoid saturation.

All observations were made with a sloan-$i'$ filter, and were reduced using standard bias, dark, and flat-fielding techniques. Relative aperture photometry was performed using AIJ \citep{Collins} on the defocused images to obtain the time-series light curve. No detrending parameters were used in the initial reductions.

During these observations, the next regular primary eclipse of the 8-day binary, occurring shortly after the second great eclipsing event, was successfully observed.

\vspace{0.2cm}

We also organized a second follow-up ground-based observing campaign about three months later to catch the {\em secondary} outer eclipses (i.e., the events when the members of the 8-day binary were eclipsed by star A in its outer orbit). We were less fortunate in observing these events than before; however, partial observations of one of the two predicted events, as well as some additional, away-from-outer-eclipse observations made it possible to further narrow the error bars on some of the orbital parameters. Furthermore, during this campaign, an additional secondary eclipse of the 8-day binary was also observed. The following observatories took apart in this second campaign.

\vspace{0.2cm}

\noindent
{\em PEST Observations}

The end of the egress phase of a secondary outer eclipse event (i.e. when one of the stars in the 8-d binary emerges from behind the disk of star A) during 2018 July 9 was observed in the $R_\mathrm{C}$ band at PEST observatory, which is a home observatory with a 12-inch Meade LX200 SCT f/10 telescope with a SBIG ST-8XME CCD camera. The observatory is owned and operated by Thiam-Guan (TG) Tan. PEST is equipped with a BVRI filter wheel, a focal reducer yielding f/5, and an Optec TCF-Si focuser controlled by the observatory computer. PEST has a $31' \times 21'$ field of view and a $1.2''$ per pixel scale. PEST is located in a suburb of the city of Perth, Western Australia. The target was also observed during the next two consecutive nights. On the night of 2018 July 10 no systematic light variations were observed, while the last observation on 2018 July 11 caught a regular secondary eclipse of the 8-day binary. These two observations, however, were not included in our analysis, since the same 8-day binary eclipse was also observed with a larger aperture telescope, which naturally produced a lightcurve with significantly less scatter (see below).

\vspace{0.2cm}

\noindent
{\em LCO Observations}

Data with the Las Cumbres Observatory (LCO; \citealp{Brown13}) were obtained from 2018 July 9 - 15. LCO a is fully robotic network of telescopes, deployed around the globe in both hemispheres\footnote{For updated information about the network see: \url{https://lco.global}}. Observing requests are entered online, including the required telescope aperture and other technical information (e.g., exposure time and band), and the scheduling software decides in which site to carry out the observation and with which telescope (many of the LCO sites contain 2-3 telescopes of the same aperture). All data obtained by LCO telescopes are reduced by an automated pipeline and made available to the users. We have carried out the photometric analysis of all LCO data obtained here using the AstroImageJ pipeline \citep{Collins}.

Data with an LCO 0.4\,m telescope in Siding Spring, Australia, were obtained on the local night of 2018 July 15. LCO 0.4m telescopes are mounted with an SBIG camera, and this data set was obtained with an exposure time of 150 s and the Pan-STARRS-w filter, while applying a telescope defocus to avoid saturation. The 2018 July 15 data set includes an ingress to a regular 8-d binary primary eclipse.  However, due to its higher photometric scatter and the missing egress phase, this resulted in an outlier ETV value and, therefore, it was omitted from the analysis.

Data with LCO 1.0m telescopes were obtained at CTIO, Chile, on the local nights of 2018 July 8 and 10, and at SAAO, South Africa, on the local nights of 2018 July 9 and 12. LCO 1.0\,m telescopes are mounted with a SINISTRO camera, where we used the SDSS-$i'$ band and an exposure time of 250 s, while defocusing the telescope by 1.0. The 2018 July 8, data includes an ingress to the first secondary outer eclipse, while the other 3 data sets obtained with the 1.0m telescopes show a flat light curve. 

Data with the LCO 2.0\,m Faulkes Telescope South (FTS) at Siding Spring, Australia, were obtained on the local nights of 2018 July 9 and 11. FTS is mounted with a Spectral camera, where we used the SDSS-$i'$ band, an exposure time of 70 second, and defocused the telescope by 1.0. The July 11 data show most of the same regular secondary eclipse of the 8-d binary, which was also observed at PEST Observatory (see above), and the July 9 data show a flat light curve.  

\begin{table*}
\caption{Mid-times of primary and secondary eclipses of the inner pair EPIC 249432662Bab}
 \label{Tab:EPIC_249432662Aab_ToM}
\begin{tabular}{@{}lrllrllrl}
\hline
BJD & Cycle  & std. dev. & BJD & Cycle  & std. dev. & BJD & Cycle  & std. dev. \\ 
$-2\,400\,000$ & no. &   \multicolumn{1}{c}{$(d)$} & $-2\,400\,000$ & no. &   \multicolumn{1}{c}{$(d)$} & $-2\,400\,000$ & no. &   \multicolumn{1}{c}{$(d)$} \\ 
\hline
57989.55756 & $-$0.5 & 0.00304 & 58026.67394 &    4.0 & 0.00089 & 58059.64242 &    8.0 & 0.00091 \\ 
57993.68484 &	 0.0 & 0.00490 & 58030.79369 &    4.5 & 0.00335 & 58063.76072 &    8.5 & 0.00312 \\ 
57997.80082 &	 0.5 & 0.00059 & 58034.92167 &    5.0 & 0.00329 & 58067.88414 &    9.0 & 0.01844 \\ 
58001.92930 &	 1.0 & 0.00597 & 58039.04110 &    5.5 & 0.00098 & 58071.99798 &    9.5 & 0.00516 \\ 
58006.04589 &	 1.5 & 0.00208 & 58043.16586 &    6.0 & 0.00234 & 58076.12270 &   10.0 & 0.00240 \\ 
58010.17639 &	 2.0 & 0.00237 & 58047.28020 &    6.5 & 0.00350 & 58207.96333 &   26.0 & 0.00113 \\ 
58014.29779 &	 2.5 & 0.00216 & 58051.40438 &    7.0 & 0.00361 & 58310.96132 &   38.5 & 0.00042 \\ 
58022.54777 &	 3.5 & 0.00106 & 58055.52069 &    7.5 & 0.01377 & 58315.06936 &   39.0 & 0.00254 \\
\hline
\end{tabular}

{\bf Notes.} Integer and half-integer cycle numbers refer to primary and secondary eclipses, respectively. Most of the eclipses (cycle nos. $-0.5 ~{\rm to} \,10.0$) were observed by {\em Kepler} spacecraft.  The last three eclipses were observed by ground-based telescopes, and the very last point was omitted from the analysis.
\end{table*}

For the combined photodynamical analysis (see Sect.\,\ref{sec:model}) the data points of each night, given in magnitudes, were converted into linear fluxes, averaged into bins with a cadence time of $\sim$$15$-min, and then normalized so that the out-of eclipse level for each night is close to unity. Then those individual data sets, which were selected for further analysis, were subdivided into two groups, according to the different filters used in the data collect. The sloan $i'$ lightcurves of the DEMONEXT and LCO 1.0 and 2.0m observations were collected into one file, while the JBO observations and the July 9 PEST observations were added to the second lightcurve file which was designated as $R_C$-band observations. 

The most problematic aspect of forming these lightcurves was in finding the correct, common, out-of eclipse flux level for the different observations. This was especially problematic for those observations where relatively little out-of eclipse data were recorded. In order to refine our initial estimated normalizations, after some preliminary photodynamical runs (see Sect.\,\ref{sec:model}), we renormalized each night's lightcurve segment via the use of our synthetic model lightcurves.  In summary, we cannot exclude the possibility of some minor systematic  effects due the inexact lightcurve normalizations. However, we expect that these systematic errors due to the uncertain out-of-eclipse-levels should be smaller than the random errors from the statistical scatter of the individual data points even after the 15-min time binning.

\section{ETV data}
\label{sec:ETV}

As usual, our first step in confirming our hierarchical triple-star hypothesis was to check whether the regular eclipses of the 8-day eclipsing binary does indeed exhibit ETVs. Therefore, we determined the mid-eclipse times of the shallow binary eclipses, and generated ETV curves. The method was the same one we used in several of our previous works, and is described in detail in \citet{borkovitsetal16}. For the $\sim80$-day-long {\em K2} observation we were able to determine the eclipse times of 10 primary and 11 secondary eclipses, which include all but one of the eclipsing events during Campaign 15. The only missing event, a primary eclipse at BJD\,2\,458\,018, occurred during the great eclipse and, therefore, it cannot be distinguished from the composite lightcurve. Later, during the two ground-based follow-up campaigns, three additional 8-day binary eclipses were also observed. The mid-eclipse times of these events were also calculated in the same manner as for the eclipses observed with {\em K2}.  Note, however, that for the very last event, due to insufficient coverage and larger photometric scatter, an outlier ETV result was obtained; therefore, this point was excluded from the subsequent analysis. All the mid-eclipse times that we obtained are tabulated in Table\,\ref{Tab:EPIC_249432662Aab_ToM}. The resultant ETV curves clearly reveal highly significant variations due to physical interactions among the stars. The combined photodynamical analysis of these ETV curves, together with the lightcurves and RV data, is described later in Sect.\ref{sec:model}. 

\section{RV Data}
\label{sec:RV}

\noindent
{\em Keck HIRES Observations} 

\vspace{3pt}

Using the Keck I telescope and HIRES instrument, we collected 4 observations of EPIC\,249432662 from 2018 Feb 1 to 2018 May 25 UT using the standard California Planet Search setup \citep{Howard10}.  With a visual magnitude of 14.9, the C2 decker ($0.87'' \times 14.0''$ was required for sky subtraction, allowing the removal of night sky emission lines and scattered moonlight that can inhibit the determination of systemic radial velocities and stellar properties. With a resulting spectral resolution of $\sim$60,000, each of the four ten-minute exposures resulted in a SNR $\sim$15 at 5000 \AA. We searched each of the four spectra for the presence of secondary spectral features due to the companion M-dwarfs and found no evidence of the companions down to a level of 1\% of the flux of the primary, and outside the separation of $\pm 10$ km s$^{-1}$ \citep{Kolbl15}. This is consistent with the expected flux from the two M-dwarf companions. Knowing that the primary star dominates the flux of the system, we calculated the systemic RV of the system using the telluric absorption lines in the HIRES red chip \citep{Chubak12} resulting in values with uncertainties of 0.2 km s$^{-1}$ (Table\,\ref{tbl:RVs}). 

\vspace{0.2cm}

\noindent
{\em McDonald Tull Spectrograph Observations} 

\vspace{3pt}

We observed EPIC\,249432662 on three different occasions with the high-resolution Tull spectrograph \citep{tull95} on the 2.7 meter telescope at McDonald Observatory in Ft.~Davis, TX.  We observed using a 1.2 arcsecond wide slit, yielding a resolving power of 60,000 over the optical band. On each night,we obtained 3-4 individual spectra back to back with 20 minute exposures to aid in cosmic-ray rejection, which we combined in post-processing to yield a single higher signal-to-noise spectrum. We bracketed each set of exposures with a calibration exposure of a ThAr arc lamp to precisely determine the spectrograph's wavelength solution. We extracted the spectra from the raw images and determined wavelength solutions using standard IRAF routines, and we measured the star's absolute radial velocity using the Kea software package \citep{endl_cochran16}. 

\vspace{0.2cm}

\begin{figure}
\begin{center}
\includegraphics[width=0.47 \textwidth]{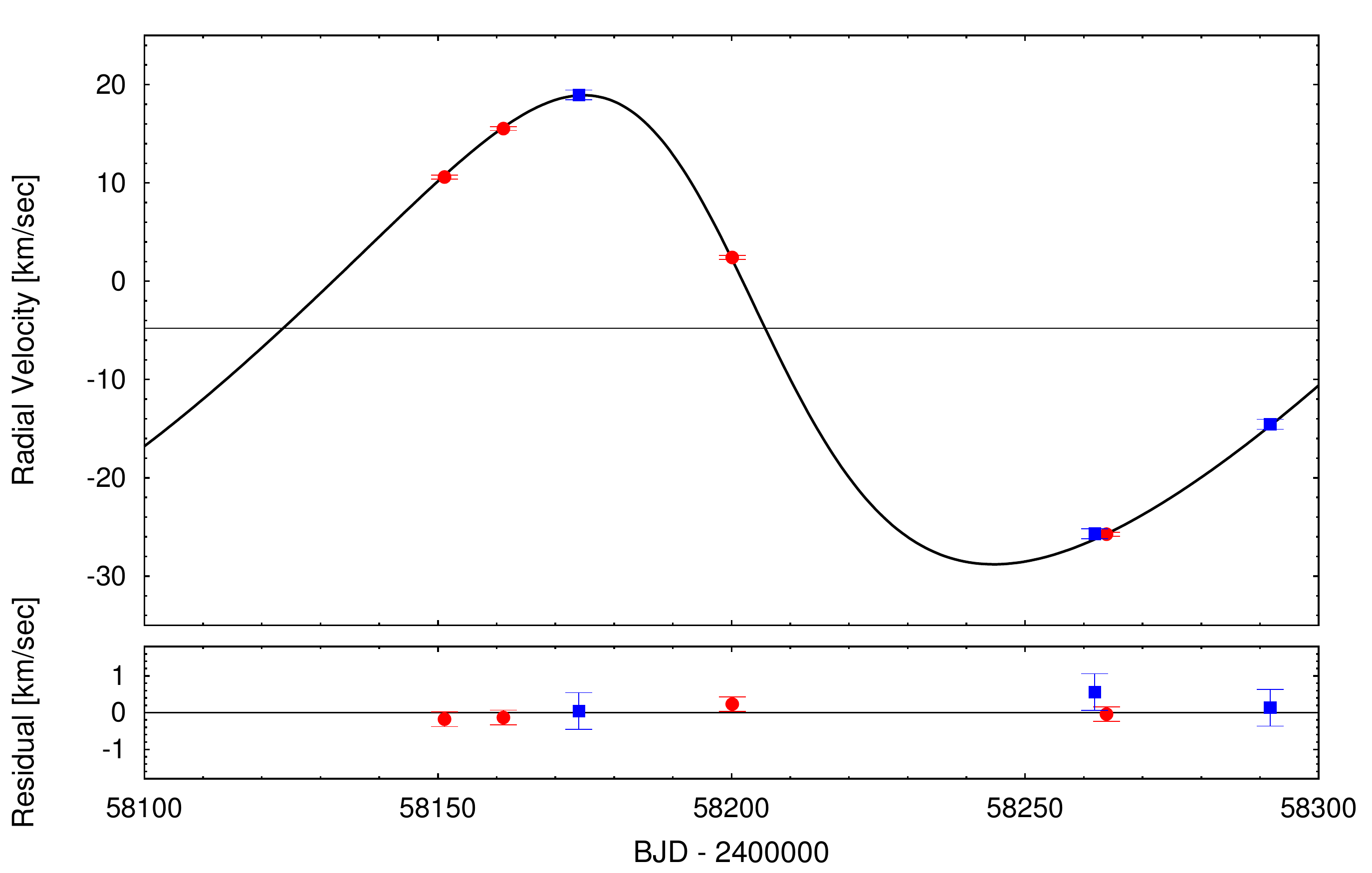}
\caption{Radial velocity measurements of the brightest, outer component of EPIC\,249432662 together with the photodynamical model RV curve (top panel), and the residuals (bottom). Red circles and blue squares denote Keck HIRES and McDonald points, respectively. The thin horizontal line in the upper panel shows the RV value at the conjunction points (i.\,e.,\, when the sum of the true anomaly and the argument of periastron of the outer orbit is equal to $\pm90\degr$). See Sect.~\ref{sec:model} for a description of the photodynamical model in which the RV points were included in the fit.} 
\label{fig:RV} 
\end{center}
\end{figure}

The seven radial velocities we were able to obtain for that outer-orbit star are shown in Fig.~\ref{fig:RV}.  The observation times and RVs with uncertainties are given quantitatively in Table\,\ref{tbl:RVs}.  There is marginally sufficient information to fit these 7 points to a general eccentric orbit, and we do not report that attempt here.  Instead, we fit these RV points simultaneously with all the other photometrically obtained data using our photodynamics code (see Sect.~\ref{sec:model}).

We determined the stellar properties of this primary star using each of the four HIRES spectra using the SpecMatch-emp routine \citep{Yee17}. The observed spectra are each shifted to the observatory rest frame, de-blazed and compared, in a $\chi^2$-squared sense to a library of previously observed HIRES spectra that span the main sequence. The best matches are determined and the values are a weighted average of the mostly closely matching spectra.  The average stellar properties resulting from analyzing all four spectra are $T_{\rm eff}= 4672\pm100$ K, $R_{\rm star}=0.77\pm0.1R_\odot$, and metallicity = $0.09\pm0.1$.  The results are given in Table \ref{tbl:RVs}.  

\begin{table} 
\centering
\caption{Radial Velocity Study}
\begin{tabular}{lc}
\hline
\hline
 & EPIC 249432662   \\
\hline
RV Measurements: &  \\
\hline
BJD-2400000 & km s$^{-1}$  \\
\hline
58151.1240$^a$  & $+10.60 \pm 0.50$  \\			
58161.1477$^a$  & $+15.50 \pm 0.50$   \\			
58173.9962$^b$  & $+18.96 \pm 1.0$  \\			
58200.1271$^a$  & $+2.42 \pm 0.50$  \\
58261.8494$^b$  & $-25.69 \pm 1.0$  \\			
58263.8553$^a$  & $-25.73  \pm 0.50$  \\
58291.7400$^b$  & $-14.55 \pm 0.84$ \\			
\hline
Spectroscopic Parameters: &  \\
\hline
$T_{\rm eff}$ [K] & $4672 \pm 100$  \\
$R$ [R$_\odot$] & $0.77 \pm 0.1$  \\
Fe/H [dex] & $0.09 \pm 0.1$   \\  
\hline  
\end{tabular} 

{\bf Notes.} $a$: Keck HIRES data; $b$: McDonald data.
\label{tbl:RVs}
\end{table}

\section{Combined photodynamical analysis}
\label{sec:model}

The compactness (i.e. the small $P_2/P_1$ ratio) of this hierarchical triple indicates that the orbital motion of the three stars is expected to be significantly non-Keplerian, even over the timescale of the presently available data.  Hence, an accurate modeling of the observed photometric and spectroscopic data, and therefore a proper determination of the system parameters, requires a complex {\em photodynamical} analysis.  This consists of a combined joint analysis of the lightcurve, ETV curve, and RV curve with a simultaneous numerical integration of the orbital motion of the three stars. 

The analysis was carried out with our own software package {\sc Lightcurvefactory} \citep{Borko13,Rappaport17,borkovitsetal18}. This code is able to emulate simultaneously the photometric lightcurve(s) of triply eclipsing triple stars (in different filter bands), the RV curves of the components (including modeling of higher order distortions of the RV curves due to e.g. Rossiter-McLaughlin effect and ellipsoidal light variations), and the ETV curves (both primary and secondary) of the inner binary. Furthermore, the motions of the three bodies, optionally, can be integrated numerically (as in the present case) or can be treated as the sum of two Keplerian motions (as is usual for hierarchical triples with negligible short-term dynamical perturbations). The built-in numerical integrator is a seventh order Runge-Kutta-Nystr\"om integrator \citep{fehlberg74}, and is identical to that which was described in \citet{borkovitsetal04}. (In Appendix\,\ref{app:numint} we also discuss  some of the practical issues regarding the use of a numerical integrator in photodynamical modeling.) Furthermore, independent of whether unperturbed Keplerian motion or numerical integration is applied, the software takes into account the light-travel time effect (`LTTE') by computing the apparent positions of the stars when light from each of them actually arrives at the Earth.  Therefore, the LTTE is inherently built into the model lightcurves.

The {\sc Lightcurvefactory} code also employs a Markov Chain Monte-Carlo (MCMC)-based parameter search, using our own implementation of the generic Metropolis-Hastings algorithm \citep[see, e.g.][]{Ford}.  Apart from the inclusion of the ETV curves, and the numerical integration of the orbital motion, the basic approach and steps of the present study are similar to those which were followed during the previous non-photodynamical analyses of two doubly eclipsing quadruple systems EPIC\,220204960 \citep[][Sect.\,7]{Rappaport17} and EPIC\,219217635 \citep[][Sect.\,7]{borkovitsetal18}. 

Here we concentrate mainly on the differences in the {\sc Lightcurvefactory} code used in this work compared to the previous studies mentioned above. The most noteworthy new feature about the present system is the existence of outer eclipses, i.e., when the inner binary occults the third star in the system or, vice versa, when the third star eclipses one or both members of the inner binary. This carries significant extra information about the geometrical configuration of the entire triple system, including both astrophysical and key orbital parameters. For example, as was shown in a number of previous studies \citep[see, e.g.][]{carteretal11,Borko13,masudaetal15}, the precise brightness variations during outer eclipses, including the timings, durations, depths, and fine structure of the eclipses, depend extraordinarily strongly on the physical dimensions of the system and, therefore, on the masses of the components. Similarly, the third-body eclipse structure depends very strongly on the orientations of the two orbits, both relative to each others and to the observer. 

Furthermore, given the compactness of this triple, the $P_2$-timescale dynamical perturbations strongly influence not only the properties of the outer eclipsing events, but also dominate the ETVs of the regular eclipses of the inner EB. This fact also offers the very good possibility of obtaining accurate orbital and dynamical parameters for this triple \citep[see][for a detailed theoretical background.]{borkovitsetal15}.

As a consequence, during our MCMC parameter search we typically jointly fit the following five data series\footnote{In some circumstances we jointly fit only a subset of these time series.}: 
\begin{itemize}
\item[(i)]{The processed, ``flattened'' {\em K2} lightcurve;}
\item[(ii-iii)]{Two sets of ($R_\mathrm{C}$- and $i'$-band) ground-based photometric observations;} 
\item[(iv)]{The RV curve of the brightest, outer component; and}
\item[(v)]{The ETV curves of the inner EB (both for primary and secondary eclipses)}.
\end{itemize}

Some of these items require some further explanation. Starting with item (i), we used two different versions of the {\em K2} lightcurve.  We made a series of MCMC runs using the complete flattened {\em K2} lightcurve (hereafter `complete' lightcurve), and another series where the out-of-eclipse sections of the lightcurve were eliminated from the fit (hereafter `eclipses-only' lightcurve). The latter results in a data train that consists only of the great eclipse itself plus a narrow window of width $\sim$0.33 days centered on each 8-d-eclipse. Dropping the out-of-eclipse sections can be easily justified by noting the spherical shape of the stars and, consequently, the lack of any measurable ellipsoidal variations (ELVs). (Note that the lack of ELVs as well as any irradiaton effect were already invoked to justify flattening the lightcurve.) There are at least two advantages in omitting the out-of-eclipse lightcurve sections. As noted, in the present long-cadence {\em K2} lightcurve each 8-day-binary eclipsing event contains only 5-6 data points, as opposed to hundreds of points in the out-of-eclipse sections. Therefore, dropping out these latter points makes the $\chi^2$-probe more sensitive to the lightcurve features during the 8-day binary eclipses and, it also saves much of the computational time. 

When we fit the `complete' {\em K2} lightcurve we did not employ any correction for the long-cadence time because of the high computational costs. However, for the fits to the `eclipses-only' {\em K2} lightcurve we corrected the model lightcurve for the $\sim$$29.4$-min long-cadence time of {\em Kepler}.\footnote{In the case of the cadence-time correction our code for each data point calculate five flux values evenly spaced within the cadence time, and then computes a net flux using Simpson's rule.} We find that our fits to the `complete' lightcurve and `eclipses-only' lightcurve result in very similar parameter values. The differences are far below the $1\sigma$ uncertainties for most of the fitted and computed parameters except for the inclination angle of the 8-day binary ($i_1$).  For $i_1$, not surprisingly, we found a bit higher values (by $\sim0\fdg1-0\fdg2$) in the cadence-corrected runs. Therefore, in the forthcoming discussion we refer to the results obtained from the `eclipses-only' runs.

Turning to item (iv), i.e. the RV curve, a visual inspection leads us to believe that there might be a few hundred m\,s$^{-1}$ offset between the Keck HIRES and the McDonald measurements.  One way to handle such discrepancies in the case of multi-site spectroscopic data is to introduce an `offset' term for {\em each} instrument as additional parameters to be fitted. In our case, however, we have only seven RV points (and two of them were taken almost at the same epoch), but there are seven parameters required for the complete solution to a purely Keplerian RV orbit (amplitude, $K$, period, $P$, eccentricity, $e$, argument of periastron, $\omega$, periastron passage, $\tau$, or their equivalents, systemic velocity, $\gamma$, and the RV offset between the two instruments). Thus, at least in terms of an analysis of the RV curve by itself, we would encounter the problem of zero degrees of freedom. Thus, instead of introducing an additional offset parameter, we constrained only a single systemic velocity ($\gamma$) parameter (see below). However we checked the effect of a potential RV offset a posteriori. This was done as follows. After obtaining a tentative solution with the joint photodynamical analysis, we calculated the averages of the RV residuals for the two sources of the RV data, and accepted their difference as a probable offset. Then we subtracted this value ($\Delta\gamma=217$\,m~s$^{-1}$) from the McDonald points and made an additional joint photodynamical MCMC run, with the original RV curve replaced with this slightly modified one.  We found from this exercise that the effect of any RV offset remains far below the $1\sigma$ parameter uncertainties. In particular, we found that the stellar masses differed by less than 1\% in either analysis. Therefore, we conclude that the presence of any small, but uncertain, RV offset has no influence on the accuracy of our parameter determination.

Regarding item (v) above, one may make the counter-argument that the accurate timings of the inner, regular eclipses are already inherent in the lightcurve analysis and, therefore, the inclusion of the ETV curves into the fitting process would be unnecessarily redundant.  While, in theory, this is evidently true, we decided to use the ETV curves for two practical reasons. First, this treatment allows us to give much higher weight to that part of the timing data which carries crucial information about the dynamics of such triple systems. By contrast, as was mentioned above, the full {\em K2} lightcurve contains only $5-6$ data points in each 8-day binary eclipsing event, compared to hundreds of points in the out-of-eclipse region. This fact makes it almost impossible to fine tune the timing data with a $\chi^2$-probe of the `complete' lightcurve fit. By contrast, the ETV curve, which is an extract of all the timing data, and contains most of the dynamical information, becomes very sensitive to even the smallest changes in the key-parameters (not just the binary period). Interestingly, we found that the same argument also remains valid for the `eclipses-only' {\em K2} lightcurve runs. In our opinion, this is so because the nearly 2-day-long great eclipse itself contains nearly the same number of data points as all of the brief 8-day-binary eclipsing events combined.  This results in an over-optimization of the great eclipse at the expense of the 8-day binary eclipses.  Second, the inclusion of the ETV curve into the photodynamical analysis has allowed us to constrain the inner orbital period during each trial step, as will be discussed below in Appendix\,\ref{app:numint}.  The practical way in which we produced the numerical model ETV data is also described in Appendix\,\ref{app:ETV}.

During our analysis we carried out almost a hundred MCMC runs, and tried several sets and combinations of adjustable parameters.  We also applied a number of physical (or technical) relations to constrain some of the parameters in order to reduce the degrees of freedom in our problem. For example, in some preliminary runs we tried to constrain the masses and/or the radii of some of the stars via empirical mass-radius-temperature relations available for main-sequence stars (e.g., \citealt{toutetal96};  \citealt{Rappaport17}; Appendix), but these runs resulted in significantly higher $\chi^2$ values and, therefore, we stopped applying such constraints. By contrast, we found it worthwhile to apply some technical (or, mathematical) constrains to the systemic radial velocity $\gamma$, the period $P_1$, and the reference primary eclipse time $(T_0)_1$ of the inner binary. In the case of $\gamma$, which is practically independent of any other parameter, its best-fit value was calculated a posteriori in each trial run with a linear regression by minimizing $\chi^2_\mathrm{RV}$ of the actual model RV curve. In the case of $P_1$ and $(T_0)_1$ the ETV model was used for the constraining process, and it is idealized and described in Appendix\,\ref{app:numint}.

For the final runs, we ended up adjusting 18 parameters, as follows: 
\begin{itemize}
\item[(i)]{Three parameters related to the remaining orbital elements of the inner binary: eccentricity ($e_1$), the phase of the secondary eclipse relative to the primary one ($\phi_\mathrm{sec,1}$) which constrains the argument of periastron ($\omega_1$), and inclination ($i_1$)\footnote{For the rigorous meaning of the orbital elements in a photodynamical problem see Appendix\,\ref{app:numint}.};}
\item[(ii)]{Six parameters related to the outer orbital elements: $P_2$, $e_2\sin\omega_2$, $e_2\cos\omega_2$, $i_2$, the time of the superior conjunction of the tertiary outer star $(T_0)_2$, and the longitude of the node of the outer orbit of the tertiary ($\Omega_2$)\footnote{The dynamical perturbations are sensitive only to the difference in the nodes ($\Delta\Omega=\Omega_2-\Omega_1$). Hence we set $\Omega_1=0\degr$, and it was not adjusted during the runs. In such a manner, adjusting $\Omega_2$ is practically equivalent to the adjustment of $\Delta\Omega$. On the other hand, however, note that due to the third-body perturbations $\Omega_1$ was also subject to low-amplitude variations during each integration run.};}
\item[(iii)]{Three mass-related parameters: the spectroscopic mass function of the outer orbit $f_2(m_\mathrm{B})$, the mass ratio of the inner orbit $q_1$, and the mass of the tertiary, which is the most massive component $m_\mathrm{A}$;}
\item[(iv)]{and, finally, six other parameters which are related (almost) exclusively to the lightcurve solutions, as follows: the fractional radii of the inner binary stars $R_\mathrm{Ba}/a_1$, $R_\mathrm{Bb}/a_1$, the physical radius of the tertiary $R_\mathrm{A}$, the temperature of the tertiary $T_\mathrm{A}$, the temperature ratio of the primary of the inner binary and the tertiary $T_\mathrm{Ba}/T_\mathrm{A}$ and, the temperature ratio of the two components of the inner binary $T_\mathrm{Bb}/T_\mathrm{Ba}$.} 
\end{itemize}  

The adjustment of $T_\mathrm{A}$ warrants some further explanation. This parameter is a natural output of the spectroscopic analysis (see Sect.\,\ref{sec:RV}), but it has only a minor influence on the lightcurve through the different relative eclipse depths in the three photometric bands.  Therefore, our original idea was to take the results of our spectroscopic analysis and then use Gaussian priors for this parameter to obtain the effective temperature of the tertiary from the complex analysis.  
However, we found that this was too constraining on $T_{\rm A}$, and in particular led to model V magnitudes that were too high and inferred distances to the source that were significantly closer than that given by Gaia.  Thus, we ultimately replaced the Gaussian prior on $T_{\rm A}$ with a uniform prior that was centered on the spectroscopic result, but the boundaries of the allowed parameter domain were expanded to somewhat beyond the $2\sigma$ uncertainties of the spectroscopic results.

Regarding the other parameters, similar to our approach with two quadruple systems \citep{Rappaport17,borkovitsetal18}, we applied a logarithmic limb-darkening law, where the coefficients were interpolated from the pre-computed passband-dependent tables in the {\sc Phoebe} software \citep{Phoebe}. The {\sc Phoebe}-based tables, in turn, were derived from the stellar atmospheric models of \citet{castellikurucz04}. Considering the gravity darkening exponents, for the nearly spherical stellar shapes in our triple, their numerical values have only a negligible influence on the ligthcurve solution. Thus, instead of using the recent results of \citealt{claretbloemen11} which tabulate stellar parameters and photometric system-dependent gravity-darkening coefficients in a three-dimensional grid, and would therefore require some further interpolation, we simply adopted a fixed value of $g=0.32$.  This is appropriate for late-type stars according to the traditional model of \citet{lucy67}.  We also found that the illumination/reradiation effect was quite negligible for all three stars; therefore, in order to save computing time, this effect was not taken into account. As a consequence of using a flattened $K2$ lightcurve (see Sect.~\ref{sec:K2}), which was assumed to be flat during the out-of-eclipse regions, we decided that in contrast to our previous work we would not take into account the Doppler-boosting effect \citep{loebgaudi03,vankerkwijketal10}.

Furthermore, in the absence of any information on the rotation properties of any of the stars, we assumed that the inner binary members rotate quasi-synchronously with their orbit.\footnote{The details of the initialization of the numerical integrator for quasi-synchronous rotation, taking into account even the likely orbital plane and stellar spin precession in the case of an inclined triple system, are described in Appendix\,\ref{app:numint}} For simplicity, for the third, outer stellar component we supposed that its rotational equatorial plane is aligned with the plane of the outer orbit, while its spin angular velocity was arbitrarily set to be ten times larger than its orbital angular velocity at periastron (which resulted in an $\sim11.5$-day rotational period). These arbitrary choices, however, have no significant influence on the lightcurve solution.

Finally, we note that due to the absence of any significant additional contaminating light in the {\em K2} aperture (see Sect.\,\ref{sec:K2}) we set the extra light parameter consistently to $\ell_\mathrm{x}=0$ for all the three lightcurves (i.e. no light contamination was considered).

\begin{table*}
\centering
 \caption{Orbital and astrophysical parameters from the joint photodynamical lightcurve, single line RV curve and ETV solution}
 \label{tab: syntheticfit}
\begin{tabular}{@{}llll}
  \hline
\multicolumn{4}{c}{orbital elements$^a$} \\
\hline
   & \multicolumn{3}{c}{subsystem} \\
   & \multicolumn{2}{c}{Ba--Bb} & A--B \\
  \hline
  $P$ [days] & \multicolumn{2}{c}{$8.1941\pm0.0007$} & $188.379\pm0.011$ \\
  $a$ [R$_\odot$] & \multicolumn{2}{c}{$16.21\pm0.16$} & $161.70\pm1.32$\\
  $e$ & \multicolumn{2}{c}{$0.0034\pm0.0019$} & $0.2212\pm0.0007$ \\
  $\omega$ [deg]& \multicolumn{2}{c}{$63.42\pm16.62$} & $243.93\pm0.66$ \\ 
  $i$ [deg] & \multicolumn{2}{c}{$89.776\pm0.198$} & $89.853\pm0.015$ \\
  $\tau$ [BJD - 2400000]&\multicolumn{2}{c}{$57988.964\pm0.381$} & $58009.854\pm0.224$ \\
  $T_0^b$ [BJD - 2400000]&\multicolumn{2}{c}{$57993.670\pm0.001$} & $58018.446\pm0.003$ \\
  $\Delta\Omega$ [deg] & \multicolumn{3}{c}{$0.155\pm0.432$} \\
  $i_\mathrm{m}$ [deg] & \multicolumn{3}{c}{$0.173\pm0.397$} \\
  \hline
  mass ratio $[q=m_2/m_1]$ & \multicolumn{2}{c}{$0.883\pm0.013$} & $1.139\pm0.054$ \\
$K_\mathrm{A}$ [km\,s$^{-1}$] & \multicolumn{2}{c}{$-$} & $23.727\pm0.230$ \\ 
  $\gamma$ [km\,s$^{-1}$] & \multicolumn{2}{c}{$-$} & $-7.113\pm0.230$ \\
  \hline  
\multicolumn{4}{c}{stellar parameters} \\
\hline
   & Ba & Bb &  A \\
  \hline
 \multicolumn{4}{c}{Relative quantities} \\
  \hline
 fractional radius [$R/a$]  & $0.0275\pm0.0012$ & $0.0233\pm0.0013 $   & $0.00442\pm0.00022 $ \\
 fractional luminosity [in {\em Kepler}-band]& $0.0455$  & $0.0281$    & $0.9264$\\
 fractional luminosity [in $R_C$-band]&        $0.0345$  & $0.0228$    & $0.9427$\\
 fractional luminosity [in $i'$-band]&         $0.0605$  & $0.0383$    & $0.9012$\\
 \hline
 \multicolumn{4}{c}{Physical Quantities} \\
  \hline 
 $m$ [M$_\odot$] & $0.452\pm0.014$ & $0.399\pm0.012$ & $0.746\pm0.040$ \\
 $R$ [R$_\odot$] & $0.445\pm0.019$ & $0.378\pm0.024$ & $0.715\pm0.030$ \\
 $T_\mathrm{eff}$ [K]& $3405\pm129$ & $3325\pm169$ & $4861\pm97$ \\
 $L_\mathrm{bol}$ [L$_\odot$] & $0.0240\pm0.0041$ & $0.0157\pm0.0038$ & $0.2562\pm0.0298$ \\
 $M_\mathrm{bol}$ & $8.79\pm0.19$ & $9.25\pm0.26$ & $6.22\pm0.13$\\
 $\log g$ [dex] & $4.80\pm0.04$ & $4.88\pm0.06$ & $4.60\pm0.04$\\
 \hline
$(M_V)_\mathrm{tot}$             &\multicolumn{3}{c}{$6.57\pm0.14$} \\
distance$^c$ [pc]                &\multicolumn{3}{c}{$424\pm32$}\\  
\hline
\end{tabular}

{\bf Notes. }{$a$: Instantaneous, osculating orbital elements, calculated for the time of the very first integration step; $b$: Superior conjunction of the primary component of the given binary. (In the case of the inner 8-day binary it is equivalent to the time of mid-primary eclipse);  c: Photometric distance, see text for details.}
\end{table*}

\begin{figure*}
\begin{center}
\includegraphics[width=0.47 \textwidth]{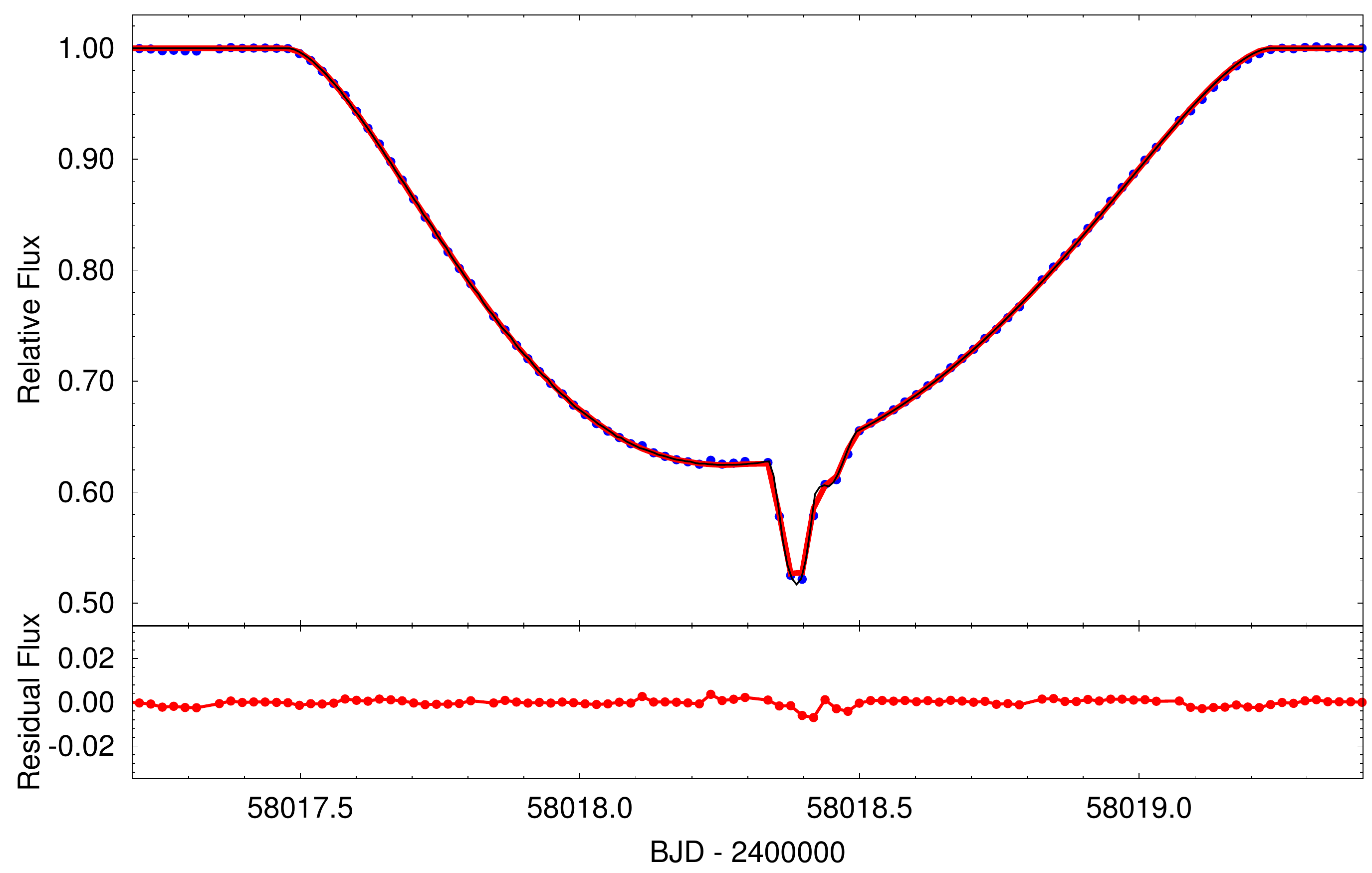}\includegraphics[width=0.47 \textwidth]{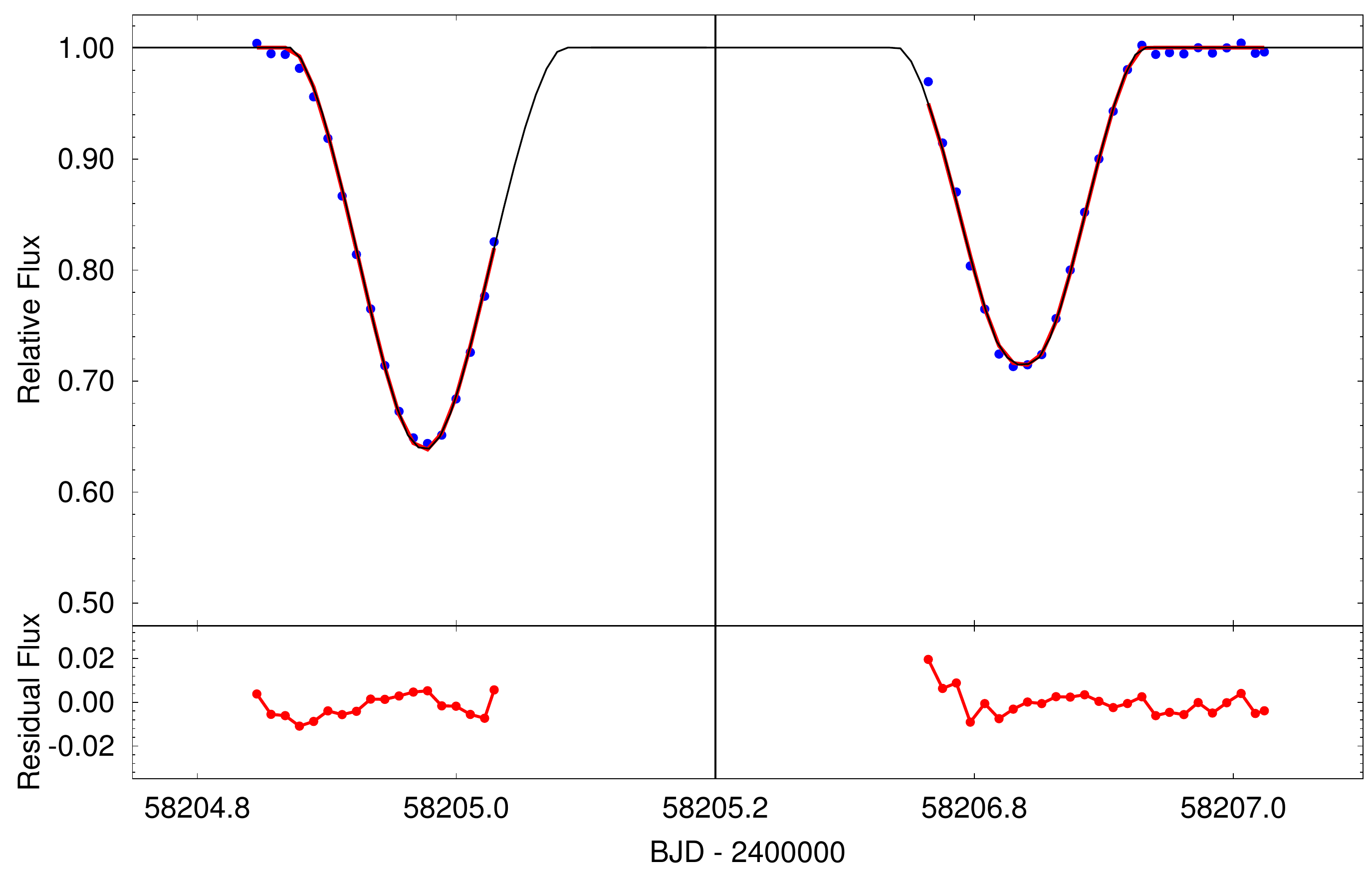}
\caption{The great eclipses of the `great eclipser'. {\em Left panel}: The outer eclipse observed by the {\em Kepler} spacecraft; {\em Right panel:} The two, separated outer eclipses, one outer orbital period (i.e., 188 days) later, observed with a ground-based telescope. The observed data are shown as blue circles. The red curve is the photodynamical model solution calculated around the time of each observation; the residuals are also shown in the bottom panels. Finally, the thin black curve represents the model solution for zero cadence time, calculated with a time step of 0.0001 of the inner binary orbital period.}
\label{fig:greateclipsesfit} 
\end{center}
\end{figure*}

\begin{figure*}
\begin{center}
\includegraphics[width=0.47 \textwidth]{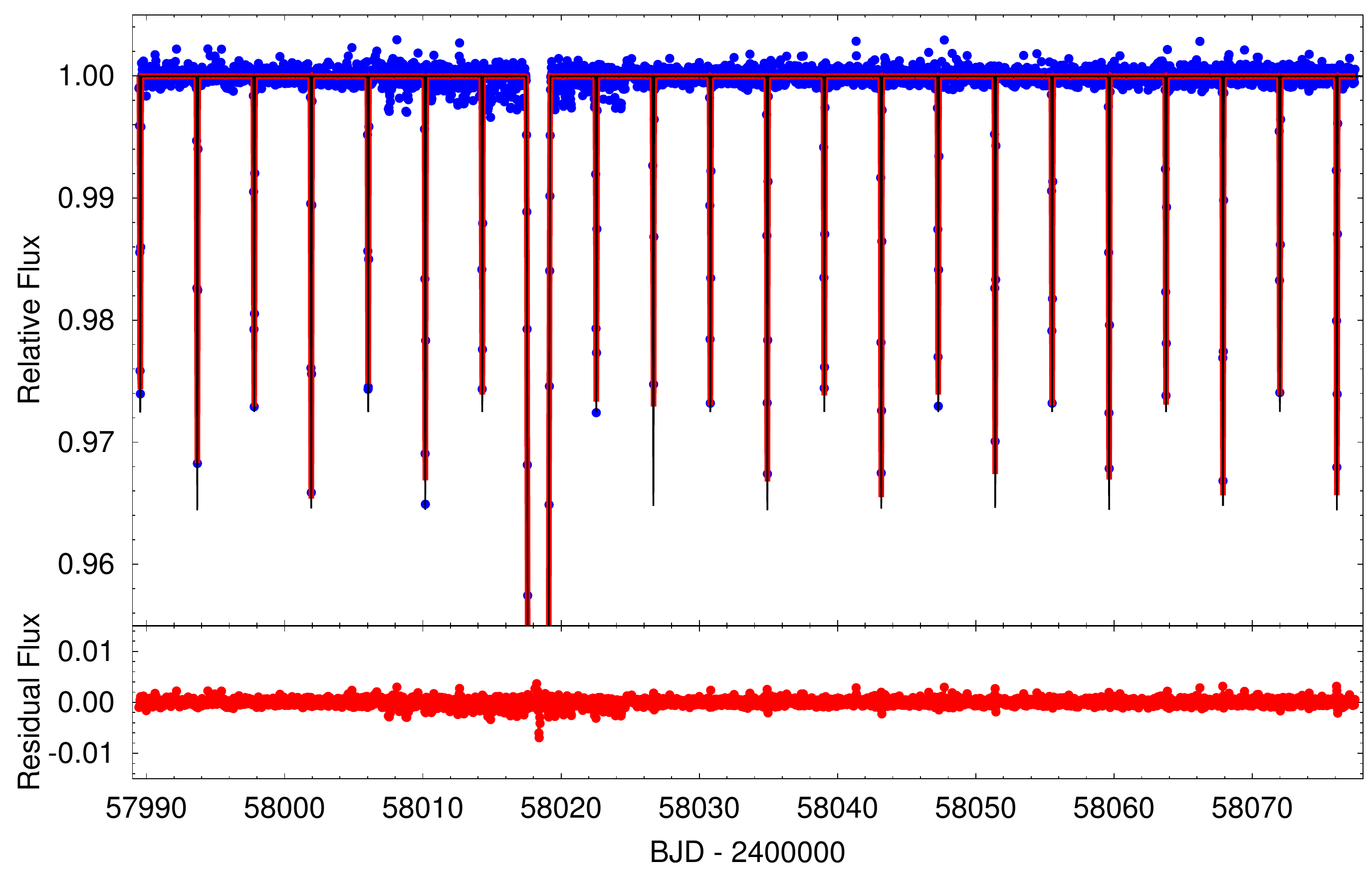}\includegraphics[width=0.47 \textwidth]{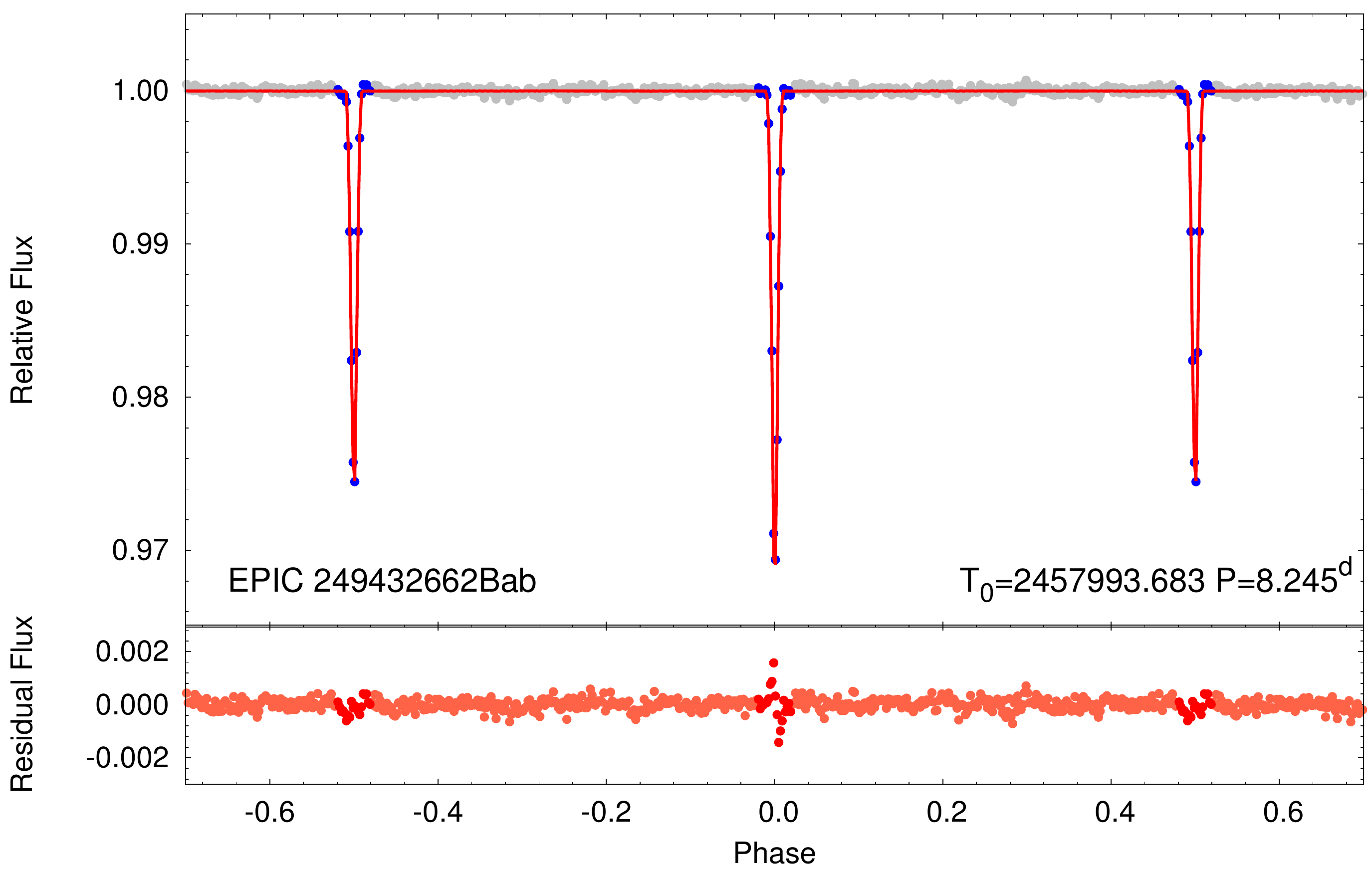}
\caption{The shallow, 8-day binary eclipses of the great eclipser. {\em Left panel}: A zoom-in on the entire {\em K2}-lightcurve, showing the regular, shallow eclipses of the inner binary pair; {\em Right panel:} The phase-folded, binned, and averaged {\em K2}-lightcurve of the inner binary after  removing the great eclipse section from the lightcurve. The observed data are shown as blue and grey circles. Note, for the `eclipses-only'-type solutions only data points within the blue phase domains were considered.  The red curve is the folded, binned and averaged lightcurve of the cadence-time corrected photodynamical model solution calculated at the time of each observation; the residuals of which are also shown in the bottom panels. Finally, the thin black curve (in the left panel) represents the model solution for zero cadence time, calculated evenly with a time step of 0.0001 of the inner orbital period. }
\label{fig:smalleclipsesfit} 
\end{center}
\end{figure*}

\begin{figure*}
\begin{center}
\includegraphics[width=0.60 \textwidth]{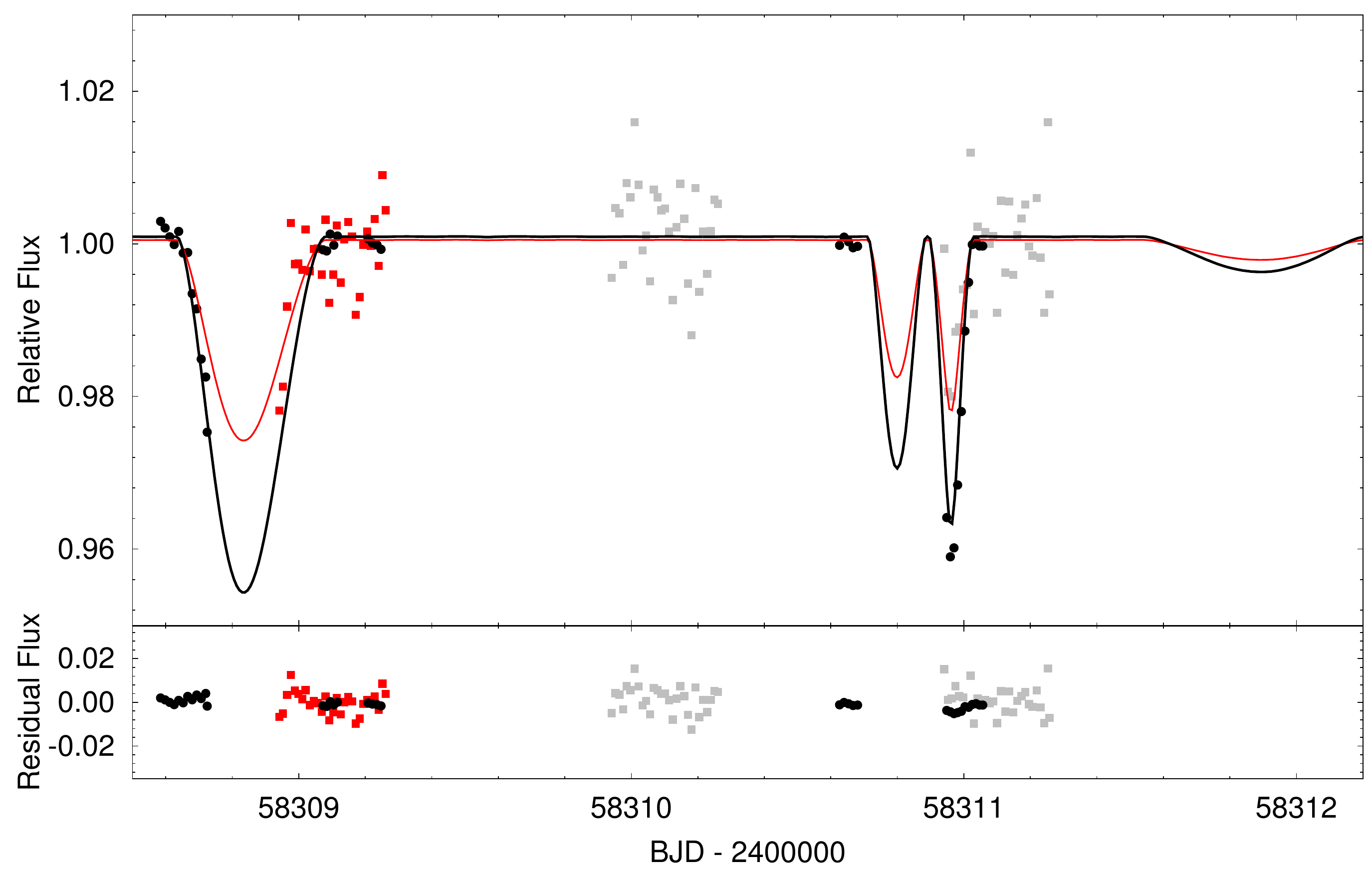}
\caption{Secondary {\it outer} (i.e., 188-day) eclipses observed during our second ground-based follow-up campaign. Three secondary outer eclipse events of the `great eclipser', for which the first was partially observed (around 58308.8), while the other two remained completely unobserved (including a shallow one near 58311.8), as well as a regular secondary eclipse of the inner (8-day) binary pair (the sharpest event at 58311) which was also partially observed. Red squares and black circles represent ground-based observations in $R_\mathrm{C}$ and $i'$ bands, respectively, while grey squares stand for further $R_\mathrm{C}$-band observations that were not used for the photodynamical analysis. Red and black lines represent the photodynamical model lightcurves in $R_\mathrm{C}$ and $i'$ bands, respectively.}
\label{fig:sececlipsesfit} 
\end{center}
\end{figure*}

\begin{figure*}
\begin{center}
\includegraphics[width=0.75 \textwidth]{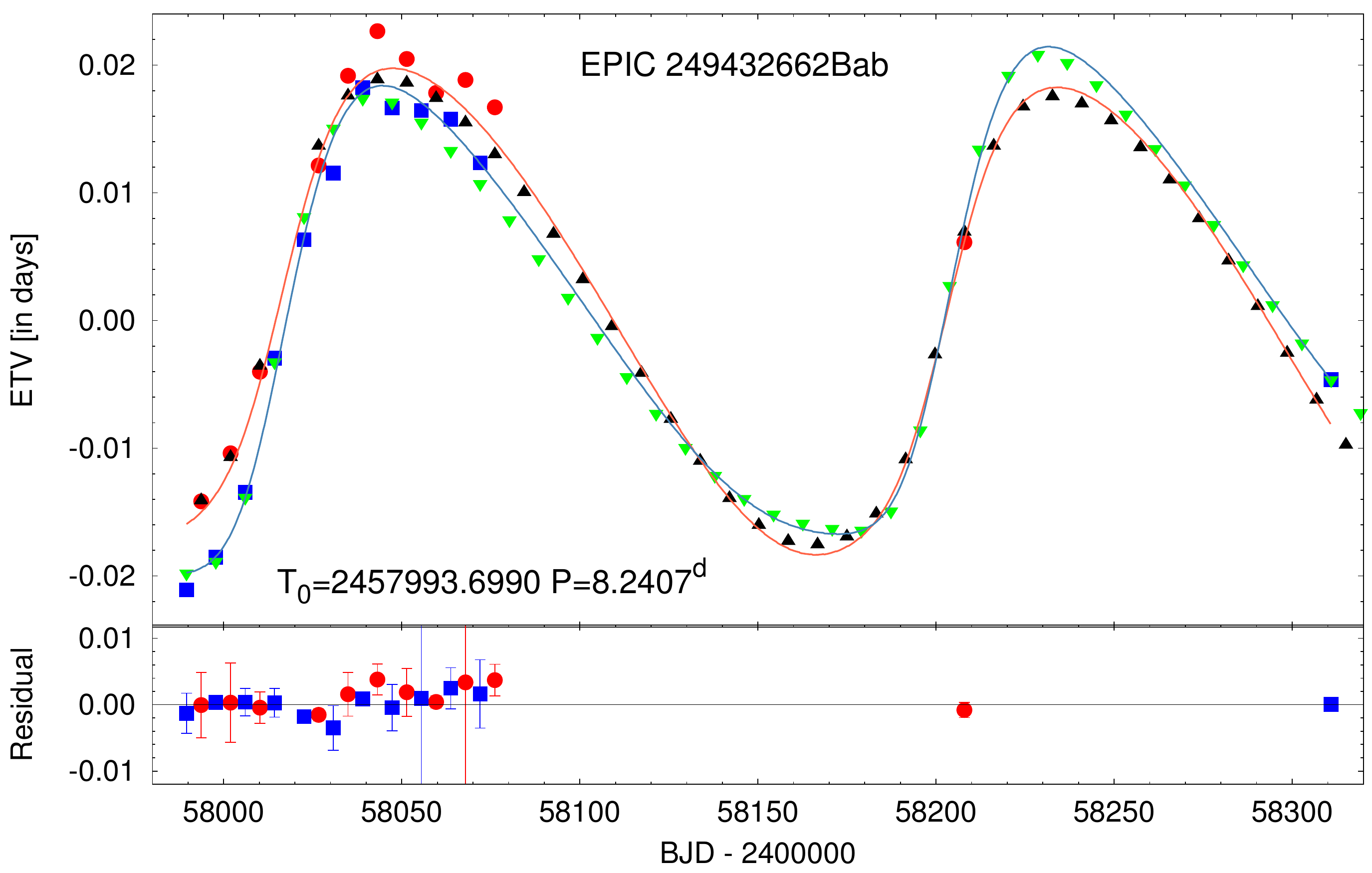}
\caption{Eclipse Timing Variations of the inner binary. {\it Upper panel:} Red circles and blue squares show the primary and secondary ETVs determined from the $K2$ and ground-based observations, while upward black and downward green triangles stand for the primary and secondary ETVs calculated from the photodynamical model, respectively.  Furthermore, the thin orange and cyan lines represent the theoretical primary and secondary ETV curves obtained from a separate analytical solution (see text for details). {\it Bottom panel:} Observed minus photodynamically modeled ETV residuals with the uncertainties of the observed ETVs. As before, red circles and blue rectangles stand for primary and secondary eclipses, respectively.}
\label{fig:ETV} 
\end{center}
\end{figure*}  

The orbital and astrophysical parameters obtained from the joint photodynamical analyses are tabulated in Table\,\ref{tab: syntheticfit}. The corresponding model lightcurves for different sections of the observed lighcurves are presented in Figs.\,\ref{fig:greateclipsesfit}, \ref{fig:smalleclipsesfit} and \ref{fig:sececlipsesfit}, while the RV-curve portion of the solution was shown previously in Fig.\,\ref{fig:RV}. Finally, the model ETV curve plotted against the observed ETVs is shown in Fig.\,\ref{fig:ETV}.

\section{Discussion of the results}
\label{sec:discussion}

\subsection{Astrophysical properties}

\begin{figure}
\begin{center}
\includegraphics[width=0.48 \textwidth]{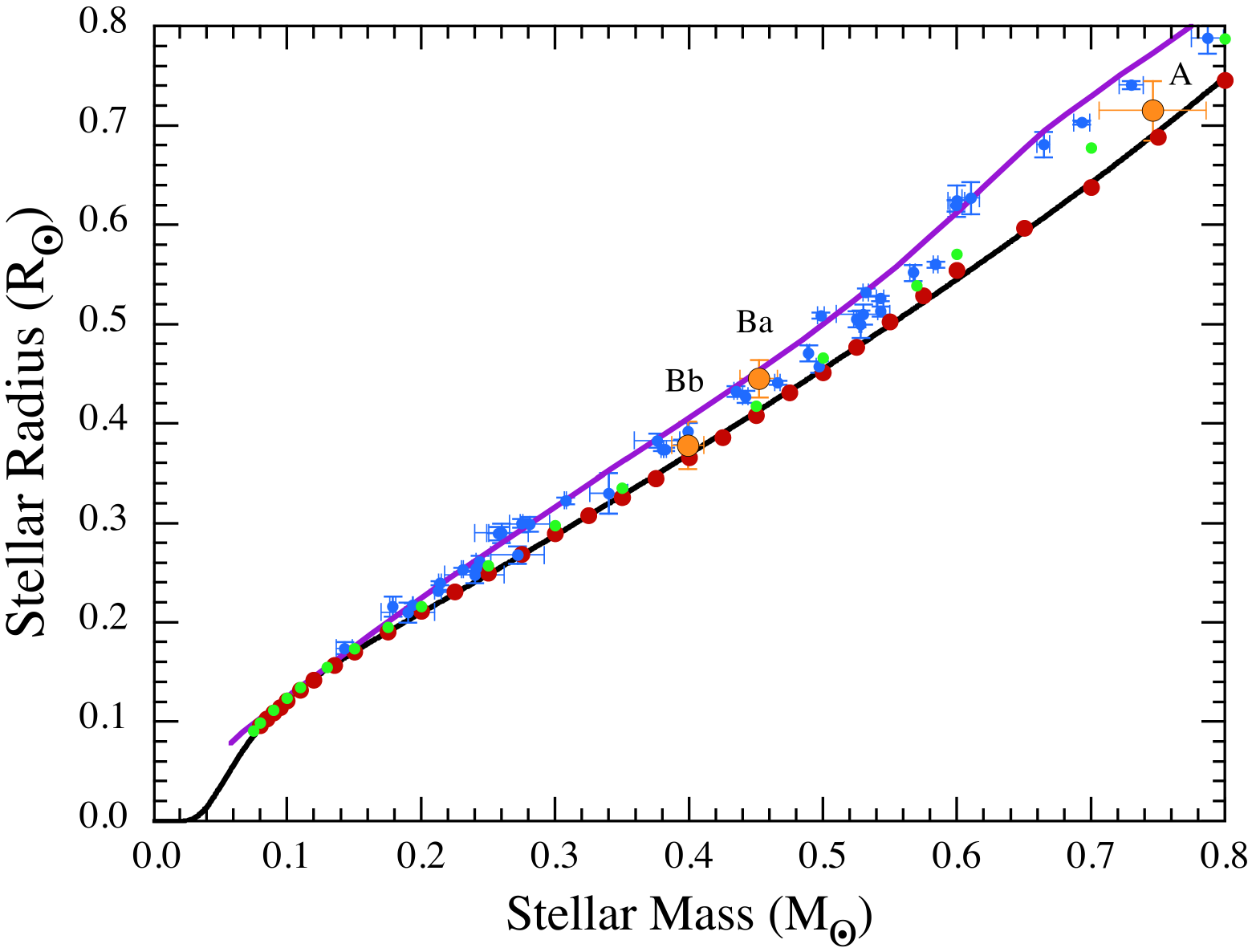}
\caption{Comparison of `great eclipser' masses and radii (large orange circles) compared with stellar radius vs.~mass relations on the lower main sequence. The red circles are models presented in \citet{Rappaport17} for solar metallicity stars. The light green circles are taken from the \citet{Baraffe98} results for similar stars. The solid black curve is the log-polynomial fit \citep[Eqn.~A1 in][]{Rappaport17} to the model points shown in red.  Blue circles with error bars are well-measured systems (see, e.g., \citealp{Cakirli,Kraus,carteretal11,Dittmann}, and references therein). The purple curve is from \citep{Stassun18} and represents the mean expected $R(M)$ value, for cool stars that are possibly thermally inflated due to their interactions in binary systems.  This figure is adapted from Fig.~16 of \citet{Rappaport17}.}
\label{fig:RM}
\end{center}
\end{figure}

Our photodynamical analysis of the available data for EPIC\,249432662, the `great eclipser', has led to a reasonably well-constrained set of system parameters (see Table \ref{tab: syntheticfit}).  Among these are the masses of the three constituent stars, which are determined well enough to make a contribution to the collection of empirically well-measured  radii and masses of stars on the lower main sequence.  We plot the $R(M)$ points for the three stars in the `great eclipser', with error bars in Fig.~\ref{fig:RM}.  Also shown in the figure are two sets of theoretical stellar models, as well as a number of well-measured stars in binary systems.  We can see that the three great eclipser stars lie somewhat above the stellar model locations, but comfortably in among the collection of other well measured stars in binary systems.  The usual explanation for the somewhat larger radii of the measured systems is thermal `inflation' due to interactions in the binary system, e.g., tidal heating \citep[see e.\,g.,][and references therein]{hanetal17}.  

A fundamental check on the system parameters that we have found can be made by computing the photometric parallax for the target, and then comparing it to the Gaia distance. To obtain this quantity using our photodynamically determined stellar radii and temperatures, we calculate the bolometric luminosities and, thereby, the absolute bolometric magnitude of each star. Then, these values are converted into absolute visual magnitudes via the formulae of \citet{flower96} and \citet{torres10}. Finally we compute the net absolute visual magnitude for the system as a whole, and obtain $M_V= 6.57 \pm 0.14$. We then utilized a web-based applet\footnote{\url{ http://argonaut.skymaps.info/query?}} to estimate $E(B-V) = 0.24 \pm 0.03$ which we translate to $A_V = 0.75 \pm 0.09$.  Thus, taking $V=15.46$ from \citet{UCAC4}, we find a distance modulus of $8.14 \pm 0.18$.  This translates to a photometric distance of $424 \pm 32$ pc, which is in quite satisfactory agreement with the Gaia-determined distance of $445 \pm 7$ pc (see Table \ref{tbl:mags}).

Note, however, that the use of a uniform prior for $T_\mathrm{A}$, instead of the spectroscopically constrained Gaussian prior, in our photodynamical analysis has resulted in an effective temperature of $T_\mathrm{A}\approx4861\pm97$\,K for the outer component, star A, which is higher by about 190\,K ($\sim$$2\sigma$) than the spectroscopically determined value of $4672 \pm 100$ K (see Sect.\,\ref{sec:RV}).  Despite the fact that, in principle, the spectroscopic temperature should be better determined than the photometric value (which is constrained indirectly through the different eclipse depths of the three separate photometric-band lightcurves), we prefer the photometric temperature, since the use of the spectroscopic value would result in a significantly lower system brightness and, therefore, a photometric distance which would be inconsistently close relative to the Gaia distance. On the other hand, however, we stress again that some caution is necessary with the currently available Gaia parallaxes. In particular, they might be rendered less accurate by systematic effects that may arise from the yet unconsidered internal 188-day motion of component A. 

\begin{figure*}
\begin{center}
\includegraphics[width=0.70 \textwidth]{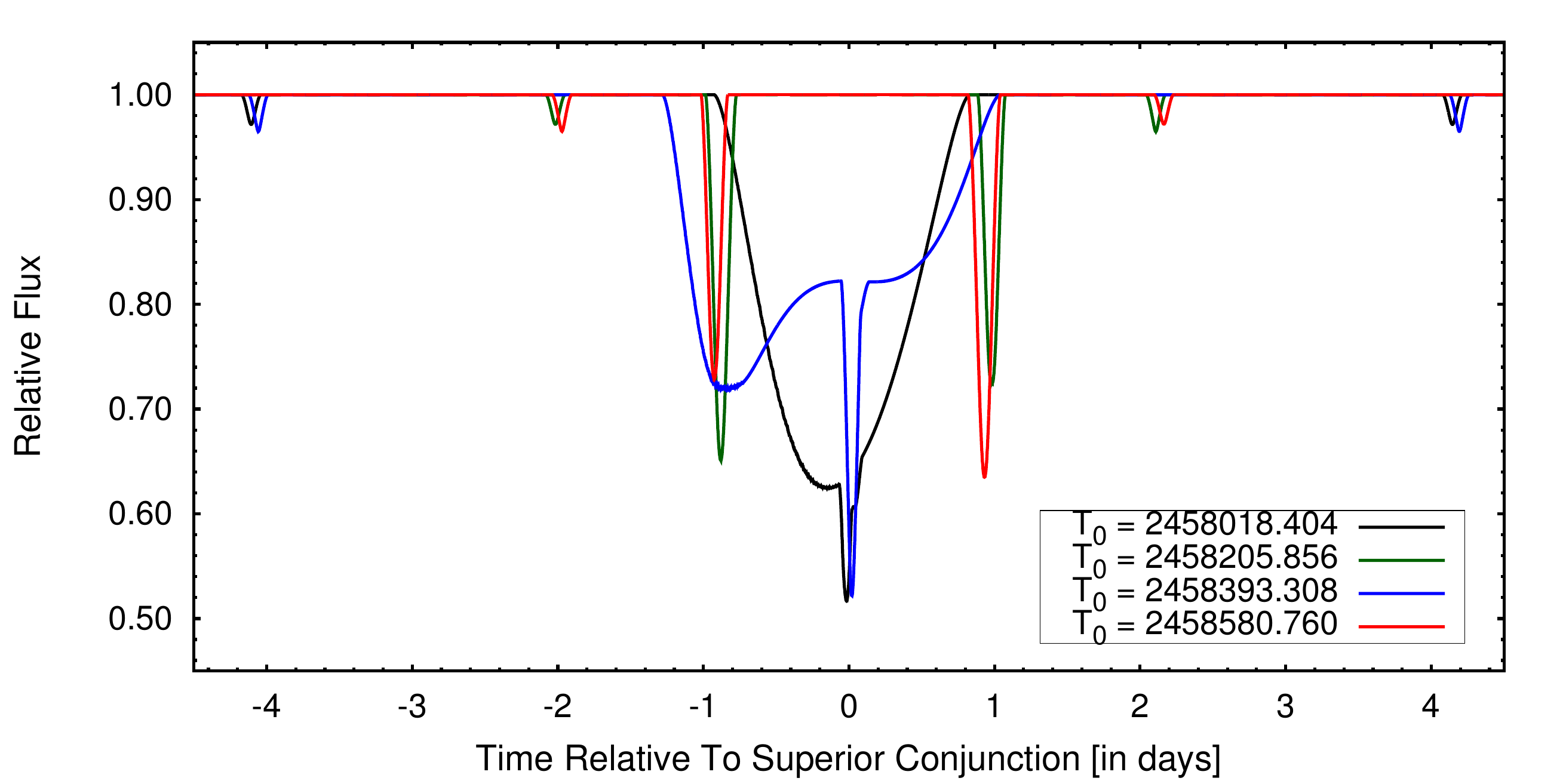}
\caption{Past and future great eclipses of EPIC\,249432662. The photodynamical model lightcurve of the {\it Kepler}-spacecraft (black) and ground-based (green) observed previous two primary {\it outer} eclipses, as well as the predicted forthcoming two events (blue and red, respectively). The closest regular, inner eclipses to the outer superior conjunctions provide a key to understanding the qualitative characteristics of the individual eclipses.  As one can see, in the black and blue cases the outer superior conjunction occurs at the time of an inner (8-day binary) eclipse. Therefore, due to the coplanar configuration, the backward component of the inner binary (i.e., the one which is eclipsed during the inner eclipse) has a very small projected velocity relative to the eclipsed third component during its transit, which results in extremely long eclipse durations. The main difference between the black and blue curves arises from the fact that the role of the two inner binary members is exchanged. A comparison of the amplitudes of the black and blue inner eclipses that are separated by $\sim$4 days from the outer eclipses, reveals that in the case of the great eclipse observed by {\em K2} the corresponding 8-day binary eclipse was a primary event, i.e. the more massive star $Ba$ was the relatively slowly moving component.  By contrast, in the forthcoming event, (which should have occurred in the fall of 2018, after the submission of this paper), shown with the blue curve, the less massive $Bb$ component with its physically faster orbital motion will move even more slowly relative to the outer stellar component, resulting in an almost 2.5-day-long great eclipse.  On the other hand, due to its smaller stellar radius, this long-lasting event will have a lower amplitude than the corresponding portion of the $K2$ event. Furthermore, as one can see, at the maximum phase of the eclipse (i.e. during the spike), when both components of the inner binary eclipse the third star, the fading will reach the same magnitude as in the case of the {\em K2} eclipse. In contrast to these two great eclipses, in the case of the green (past) and red (future) events, the outer superior conjunction occurs near to the quadrature of the inner binary and, therefore, the two binary members eclipse the third star separately, and relatively quickly. Note, that the roles of the two inner (8-day) binary members are also reversed for these two events.}
\label{fig:greateclipses_past_future} 
\end{center}
\end{figure*}

\subsection{Orbital/dynamical properties}

Our results reveal an extremely flat triple system. The mutual inclination was found to be $i_\mathrm{mut}=0\fdg17\pm0\fdg40$. Note, that this narrow error bar is actually quite realistic.  Our numerical integrations (in agreement with analytic perturbation theory) reveal that even a mutual inclination of $2\degr-3\degr$ would result in the rapid precession of the orbital planes, leading to significant variations in the eclipse depths of the inner binary even within the few-month-long observed datasets. Note that our numerical integration predicts a variation in the inclination angle of the 8-day binary's orbital plane of $\Delta{i_1}\approx-0\fdg05$ over the next 2 years.

In spite of the very good fits we obtained, we also investigated the possibility of a flat but retrograde (i.\,e., $i_\mathrm{mut}\sim$$180\degr$) scenario for the inner vs.~outer orbits. For this scenario we departed from the accepted prograde solution, but (i) added $180\degr$ to $\Omega_2$ transposing the ascending and descending nodes of the outer orbit, (ii) flagged the regular inner  binary eclipse at $T_0\simeq2457993.6848$ as being a secondary eclipse instead of primary one, thereby shifting the positions of the binary stars on their close orbit by $\sim$$180\degr$ and consequently, (iii) changed the initial phase of the secondary binary eclipse relative to the primary one by $1-\Delta\phi$ which essentially resulted in a $\sim$$180\degr$ difference in $\omega_1$.  Applying these settings, we obtained a flat, but retrograde triple system for which the lightcurve reproduced the outer eclipse patterns rather well. On the other hand, due to the phase shift of the two inner binary stars the depth ratio of the primary and secondary eclipses reversed, resulting in {\it dis}agreement with the observed lightcurve. Furthermore, for the dependence of the dynamical ETV-pattern on the observable argument of periastron ($\omega_1$), the simulated ETV curve in the retrograde scenario also departed strongly from the observed one\footnote{The usefulness of dynamical ETV curves in deciding between prograde and retrograde solutions was also noted recently by \citet{nemravovaetal16} and \citet{broz17}}. (Note, since none of these input-parameter changes had any effect on the radial velocity curve of the outer component, the latter was indifferent in this sense.)  Finally, we made some MCMC runs initiating our triple system with this parameter set, but none of these runs resulted in a suitable retrograde solution. Therefore, we can conclude with great confidence, that EPIC\,249432662 is a flat and prograde triple.

As far as we know, there is only one other known hierarchical triple star system which has been found to be so extremely flat as the `great eclipser'.  This is the triply eclipsing system HD\,181068 (KIC\,05952403) in which, similar to the present work, it was the analysis of the triply eclipsing lightcurve  that led to a similar conclusion \citep{Borko13}.  On the other hand, however, there is a significant difference between the two systems. The outer component of HD\,181068 is an evolved red giant star with a fractional radius (i.e., $R_{\rm A}/a_2$) that is larger by a factor of 25 than that in the present system ($\approx 0.138$ vs. $\approx 0.005$, respectively).  Therefore, in the HD\,181068 system tidal effects should have played a significant role in the dynamical evolution of the orbits, while this is not the case in the present triple. Interestingly, the shortest known outer-period {\em Kepler}-triple, KOI-126 (KIC\,05897826) which, similar to EPIC\,249432662, is formed by three main-sequence stars (the inner binary members are very low mass M-stars), has a significant non-zero mutual inclination ($i_\mathrm{mut}\sim8\degr$, \citealp{carteretal11}), while the similarly short outer period triple HD\,144548 is probably flat ($i_\mathrm{mut}<2\degr$), though this latter result was inferred only indirectly from the radial velocity analysis, instead of a dynamical study \citep{alonsoetal15}. 

This flatness of EPIC\,249432662, together with the very similar (and close to unity) inner and outer mass ratios, may imply that all three stellar components were formed sequentially from the same viscous disk. This formation scenario which is an extension of the equal-mass, sequential binary formation mechanism is described briefly in \citet{tokovinin18}.

Regarding the present-day dynamical behavior of our triple system, the outer-to-inner period ratio of $P_2/P_1$ is $\approx22.8$, making this triple a member of the exclusive group of 15 most compact triple stellar systems discovered amongst the {\em Kepler} EBs.  As a consequence, the orbital elements are subject to significant cyclic variations due to third-body perturbations which are readily observable even during the half-year-long timescale of the outer orbital period. The best example is the quasi-sinusoidal, $\sim$$0.03$-day full-amplitude ETV curve (see Fig.\,\ref{fig:ETV}). The ratio of the dynamical ETV amplitude relative to the light-travel-time contribution is $\mathcal{A}_\mathrm{dyn}/\mathcal{A}_\mathrm{LTTE}\sim7.7$. 

As a double check, we carried out an independent analysis of the ETV curves using our {\em analytic} ETV curve emulator and fitting software package described in \citet{borkovitsetal15,borkovitsetal16}. The mathematical model was identical with the previously described one; however, for solving the inverse problem, instead of the previously employed non-linear Levenberg-Marquardt algorithm, we also used the same MCMC algorithm which was applied in our photodynamical code. For this parameter search we fixed the mutual inclination at $i_\mathrm{m}\equiv0$ and, mode `AP3' was used for modeling the dynamically forced apsidal motion of the inner binary, i.e. the period of the apsidal motion was not an additional free parameter, but rather was calculated from the formulae of the analytical perturbation theory (in first order), as was discussed in Appendix\,C of \citet{borkovitsetal15}. The orbital elements and other parameters obtained from this analytic solution are tabulated in Table\,\ref{tab:numvsanal}. For an easy comparison to the photodynamical results, we also tabulate the averages over one binary period of the same orbital elements (calculated for the very first numerically integrated complete orbit of the inner binary, see Appendix\,\ref{app:numint}). Furthermore, the analytical ETV solution is also plotted (as thin orange and cyan lines) in Fig.~\ref{fig:ETV}.

\begin{table}
\centering
 \caption{Comparison of the results of the purely analytical ETV solution and the photodynamical analysis}
 \label{tab:numvsanal}
\begin{tabular}{@{}lll}
  \hline
Parameter & analytic ETV & photodynamical \\
\hline
$P_1$ [days]     & $8.2304\pm0.0005$ & 8.2344 \\
$a_1$ [$R_\odot$] & $13.71\pm1.08$    & 16.26\\
$e_1$            & $0.0055\pm0.0007$ & 0.0084\\
$\omega_1$ [deg] & $83.01\pm0.73$ & 84.74\\
$i_1$ [deg]      & 89.83 & 89.78 \\
$\Omega_1$ [deg] & 0.00 & 0.00 \\
$\tau_1$ [BJD - 2400000]& $57997.646\pm0.017$&\\
$P_2$ [days]     & $188.41\pm0.18$ & 188.32 \\
$a_2$ [$R_\odot$] & $153.70\pm4.94$ & 161.66\\
$e_2$            & $0.172\pm0.019$ & 0.221\\
$\omega_2$ [deg] & $63.65\pm6.04$ & 64.06\\
$i_2$ [deg]      & 89.83 & 89.85 \\
$\Omega_2$ [deg] & 0.00 & 0.08 \\
$\tau_2$ [BJD - 2400000]& $58010.1\pm4.5$ &\\
$(T_0)_2$ [BJD - 2400000]& $58019.8\pm5.7$ & 58018.4\\
\hline
$P_\mathrm{apse}$ [years] & $33\pm3$ & 23.6\\
\hline
$m_\mathrm{A}$ [$M_\odot$] & $0.86\pm0.05$ & $0.75\pm0.04$\\ 
$m_\mathrm{B}$ [$M_\odot$] & $0.51\pm0.12$ & $0.85\pm0.02$\\
\hline 
\end{tabular} 
\end{table}

We have compared the complex lightcurve+ETV+RV photodynamical results with those obtained from fits using purely analytical expressions for the ETVs. Apart from the masses, and especially the reversal of the outer mass ratio ($M_A/M_B$) in the analytic ETV case, the orbital elements we obtain nicely match each other. We stress the very solid agreement in the times of the superior conjunction of the outer orbit $(T_0)_2$ which is very strongly constrained in the joint photodynamical solution through the location and shape of the $K2$ observed outer eclipse. By contrast, in the case of a pure ETV solution, there is no strong direct constraint on this parameter, and its value can be inferred only indirectly, and with much lesser accuracy than the orbital elements obtained from the complete photodynamical solution. These results provide further, a posteriori, confirmation of the robustness of the analytical method described in \citet{borkovitsetal15}.

Numerical integrations, as well as analytical computations further reveal, however, directly nonobservable, cyclic variations on both the timescales of the inner and outer orbital periods.  These are nicely illustrated below in Fig.\,\ref{fig:orbelements_numint} of Appendix\,\ref{app:numint}. (These short timescale variations need some care in the strict interpretation of the meaning of the orbital elements tabulated in Table\,\ref{tab: syntheticfit}, as is discussed in Appendix\,\ref{app:numint}.) The third class of periodic perturbations in hierarchical triple systems are the so-called `apse-node' timescale perturbations, which have the longest period and highest amplitude \citep[see, e.\,g.][]{brown36}. These include the apsidal motion of both the inner and the outer orbits, which will already be observable in EPIC\,249432662 in the near future with ground-based follow up observations. Note, Fig.\,\ref{fig:ETV} nicely illustrates that, according to our solution, the phase displacements of the secondary eclipses of the 8-day binary relative to the primary eclipses had reversed their signs by the time of the latest outer primary eclipsing events around JD\,2458206 (i.e., for now, the primary ETV curve is running below the secondary one). This a direct consequence of the dynamical apsidal motion of the inner binary, and this reversal corresponds to that time domain when the 1-orbit-averaged $\omega_1$ becomes greater than 90\degr (see the upper right panel in Fig.\,\ref{fig:orbelements_numint}). Unfortunately, in terms of an observational verification of this model prediction, the times of only one primary and one widely separated secondary eclipse of the 8-day binary have been observed since the K2 observation. These are inconclusive as to this reversal due to the large time lapse between them. A quick and certain verification of such a reversal of the timing of the primary and secondary eclipses of the 8-day binary would require observations of a few consecutive (or nearly consecutive) primary and secondary eclipses. In order to help prepare for the encouraged future follow-up observations we tabulate the predicted times of minima for the next one year in Table~\ref{tab:inner_eclipse_predict}.

\begin{table}
\caption{Predicted inner (8-d) eclipse times for EPIC\,249432662}
 \label{tab:inner_eclipse_predict}
\begin{tabular}{@{}lclclc}
\hline
BJD & type & BJD & type & BJD & type \\ 
$-2\,400\,000$  & & $-2\,400\,000$ & & $-2\,400\,000$ &  \\ 
\hline
58422.2314    & p &  58549.9362  & s &  58677.6706   & p \\
58426.3588    & s &  58554.0472  & p &  58681.8006   & s \\
58430.4714    & p &  58558.1773  & s &  58685.9076   & p \\
58434.5977    & s &  58562.2907  & p &  58690.0385   & s \\
58438.7106    & p &  58566.4212  & s &  58694.1449   & p \\
58442.8359    & s &  58570.5379  & p &  58698.2768   & s \\
58446.9492    & p &  58574.6708  & s &  58702.3825   & p \\
58451.0739    & s &  58578.7883  & p &  58706.5156   & s \\
58455.1872    & p &  58582.9234  & s &  58710.6208   & p \\
58459.3116    & s &  58587.0377  & p &  58714.7548   & s \\
58463.4249    & p &  58591.1726  & s &  58718.8597   & p \\
58467.5492    & s &  58595.2831  & p &  58722.9947   & s \\
58471.6622    & p &  58599.4170  & s &  58727.0995   & p \\
58475.7868    & s &  58603.5255  & p &  58731.2350   & s \\
58479.8993    & p &  58607.6582  & s &  58735.3404   & p \\
58484.0244    & s &  58611.7662  & p &  58739.4754   & s \\
58488.1364    & p &  58615.8977  & s &  58743.5824   & p \\
58492.2621    & s &  58620.0059  & p &  58747.7169   & s \\
58496.3734    & p &  58624.1365  & s &  58751.8267   & p \\
58500.4999    & s &  58628.2450  & p &  58755.9619   & s \\
58504.6106    & p &  58632.3747  & s &  58760.0750   & p \\
58508.7382    & s &  58636.4835  & p &  58764.2127   & s \\
58512.8483    & p &  58640.6126  & s &  58772.4648   & s \\
58516.9767    & s &  58644.7215  & p &  58776.5742   & p \\
58521.0863    & p &  58648.8502  & s &  58780.7129   & s \\
58525.2160    & s &  58652.9591  & p &  58784.8187   & p \\
58529.3251    & p &  58657.0879  & s &  58788.9565   & s \\
58533.4557    & s &  58661.1964  & p &  58793.0606   & p \\
58537.5649    & p &  58665.3254  & s &  58797.1972   & s \\
58541.6959    & s &  58669.4335  & p &  58801.3010   & p \\
58545.8055    & p &  58673.5629  & s &  58805.4365   & s \\
\hline
\end{tabular}

{\bf Notes.} Times are given in BJD$-2\,400\,000$.  $p$ and $s$ refer to primary and secondary eclipses, respectively.
\end{table}

Regarding the forthcoming outer eclipses, according to our model, the 2018 fall event should have been the longest great eclipse at least over the following decade (see blue curve in Fig.\,\ref{fig:greateclipses_past_future}).  As one can see, this outer primary eclipse was expected to have had an extremely long duration of almost 2.3 days between 2018 Sept 30, and 2018 Oct 2. This epoch, however, was unfortunately unfavourable for ground-based follow up observations due to the source's proximity to the Sun. In Fig.\,\ref{fig:greateclipses_past_future} we also plot the previous two great (primary) eclipse events (black and green curves) together with the next upcoming one (red curve). For the convenience of observers, we list the predicted time intervals for the next few outer primary and secondary eclipses in Table\,\ref{tab:outer_eclipse_predict}. 

\begin{table}
\caption{Predicted outer eclipse times over the next 1\,000 days for EPIC\,249432662}
 \label{tab:outer_eclipse_predict}
\begin{tabular}{@{}llr}
\hline
First event & Last event &  type \\ 
\hline
58392.0 & 58394.3 & p \\
58496.9 & 58499.4 & s \\
58579.7 & 58581.8 & p \\
58683.0 & 58685.5 & s \\
58767.2 & 58768.7 & p \\
58871.5 & 58873.8 & s \\
58954.7 & 58956.8 & p \\
59057.7 & 59060.2 & s \\
59141.8 & 59143.4 & p \\
59244.1 & 59248.4 & s \\
59329.6 & 59331.6 & p \\
59432.3 & 49434.8 & s \\
59516.9 & 59518.3 & p \\
\hline
\end{tabular}

{\bf Notes.} Times are given in BJD$-2\,400\,000$, and stand for the beginning of the first fading and the end of the last fading; $p$ and $s$ refer to the primary and secondary outer (188-d) eclipses, respectively.
\end{table}

In order to test the long-term stability of this triple, we carried out 1-million-year-long numerical integrations with the same integrator that was also used during the photodynamical analysis. Irrespective of whether we used a simple three-body point-mass model, or included tidal effects, we did not detect any secular variations in the orbital elements.  Therefore, we can conclude that, not surprisingly, the present configuration of EPIC\,249432662 is stable over the nuclear timescale of star A.

\section{Summary and Conclusions}
\label{sec:conclusion}

In this work we report the discovery of a remarkable triple star system with an inner 8-day M-star binary orbiting a K star in a 188-day mildly eccentric orbit.  Both orbital planes are nearly perfectly aligned with us the observers.  The motion of one of the stars in the 8-day binary can conspire to nearly cancel its relative motion across the outer K star during opposition, thereby producing very long eclipses (up to 2 days).  

We have carried out a comprehensive photodynamical analysis of the photometric, RV, and ETV data sets.  This analysis has led to a good assessment of all the system parameters (see Table \ref{tab: syntheticfit}), including the three constituent masses, orbital parameters, and the conclusion that the two orbital planes are flat to within $\sim$$0.2^\circ$. We also checked the more quantitative photodynamical solution against fits to analytic approximations to the ETV curves in order to `calibrate' how well they work.

The masses of the two M stars in the 8-day binary are determined with about 5\% precision (1.5 $\sigma$), which is sufficient to make them interesting comparison stars to test against stellar evolution models of lower-main sequence stars.  As with the relatively few previously well determined systems, the radii are somewhat, but significantly, larger than models from the latest stellar evolution codes indicate.

The system has a substantially low ratio of outer to inner periods.  This leads to a number of different dynamical interactions which are described in the text.  Perhaps the most observationally dramatic of these is the large-amplitude ETV curve which we measure, and model both analytically and with a direct 3-body integration.

The accuracy with which the system parameters are known can be improved by future observations.  We have demonstrated that small telescopes used by amateurs can readily observe the deep and broad outer 3rd-body eclipses, as well as even the 8-day binary eclipses.  In this regard, we provide a list of times for future outer eclipses as well as eclipses of the 8-day binary.  We encourage such observations over the next couple of observing seasons.
According to our projected orbital solution, the very next upcoming outer eclipse is expected to be the longest in duration, and the deepest in amplitude, for at least the forthcoming decade.

\section*{Acknowledgements}

T.\,B. acknowledges the financial support of the Hungarian National Research, Development and Innovation Office -- NKFIH Grant OTKA K-113117.  We thank Bruce Gary for reducing the data from the JBO observations. M.\,H.\,K., M.\,R.\,O., H.\,M.\,S.~and I.\,T.~acknowledge Allan R. Schmitt for making his lightcurve examining software {\tt LcTools} freely available. We are thankful for the fruitful discussions with Sz.\,Csizmadia (DLR, Berlin) and I.\,B.\,B\'\i r\'o (Baja Astronomical Observatory, Hungary). W.\,D.\,C.~and M.\,E.~acknowledge support from NASA grants NNX16AJ11G and 80NSSC18K0447 to The University of Texas at Austin. This work was performed in part under contract with the California Institute of Technology/Jet Propulsion Laboratory funded by NASA through the Sagan Fellowship Program executed by the NASA Exoplanet Science Institute. This paper includes data collected by the {\em K2} mission. Funding for the {\em K2} mission is provided by the NASA Science Mission directorate. Some of the data presented in this paper were obtained from the Mikulski Archive for Space Telescopes (MAST). STScI is operated by the Association of Universities for Research in Astronomy, Inc., under NASA contract NAS5-26555. Support for MAST for non-HST data is provided by the NASA Office of Space Science via grant NNX09AF08G and by other grants and contracts. The authors wish to extend special thanks to those of Hawai''ian ancestry on whose sacred mountain of Maunakea we are privileged to be guests. Without their generous hospitality, the Keck observations presented herein would not have been possible.




\begin{thebibliography}{99}


\bibitem[Alonso et al.(2015)]{alonsoetal15} Alonso, R., Deeg, H. J., Hoyer, S., Lodieu, N., Palle, E., Sanchis-Ojeda, R., 2015, \aap, 584, L8

\bibitem[Armstrong et al.(2012)]{armstrongetal12} Armstrong, D., Pollacco, D, Watson, C. A. et al., 2012, \aap, 545, L4


\bibitem[Auvergne et al.(2009)]{auvergneetal09} Auvergne, M., Bodin, P., Boisnard, L. et al., \aap, 506, A411

\bibitem[Baraffe et al.(1998)]{Baraffe98} Baraffe, I., Chabrier, G., Allard, F., \& Hauschildt, P.H. 1998, A\&A, 337, 403

\bibitem[Barentsen \& Cardoso (2018)]{kadenza} Barensten, G., \& Cardosos, J.V.d.M. 2018, Astrophysics Source Code Library, ascl.soft03005B

\bibitem[Borkovits et al.(2002)]{borkovitsetal02} Borkovits, T., Csizmadia, Sz., Heged\"us, T., B\'\i r\'o, I. B., S\'andor, Zs., \& Opitz, A., 2002, \aap, 392, 895

\bibitem[Borkovits et al.(2004)]{borkovitsetal04} Borkovits, T., Forg\'acs-Dajka, E., \& Reg\'aly, Zs., 2004, \aap, 426, 951

\bibitem[Borkovits et al.(2011)]{borkovitsetal11} Borkovits, T., Csizmadia, Sz., Forg\'acs-Dajka, E., \& Heged\"us, T., 2011, \aap, 528, A53

\bibitem[Borkovits et al.(2013)]{Borko13} Borkovits, T., Derekas, A., Kiss, L.~L., Kir\'aly, A., Forg\'acs-Dajka, E., B\'{\i}r\'o, I.~B., Bedding, T.~R., Bryson, S.~T., Huber, D., \& Szab\'o, R., 2013, \mnras, 428, 1656

\bibitem[Borkovits et al.(2015)]{borkovitsetal15} Borkovits, T., Rappaport, S., Hajdu, T., Sztakovics, J., 2015, \mnras, 448, 946 

\bibitem[Borkovits et al.(2016)]{borkovitsetal16} Borkovits, T., Hajdu, T., Sztakovics, J., Rappaport, S., Levine, A., B\'\i r\'o, I. B., Klagyivik, P. 2016, \mnras, 455, 4136 

\bibitem[Borkovits et al.(2018)]{borkovitsetal18} Borkovits, T., Albrecht, S., Rappaport, S., et al., 2018, \mnras, 478, 5135

\bibitem[Borucki et al.(2010)]{boruckietal10} Borucki, W. J., Koch, D., Basri, G. et al., 2010, Science, 327, 977

\bibitem[Bro\v z(2017)]{broz17} Bro\v z, M., 2017, \apjs, 230, 19

\bibitem[Brown(1936)]{brown36} Brown, E. W., 1936, \mnras, 97, 62

\bibitem[Brown et al.(2013)]{Brown13} Brown, T. M., Baliber, N., Bianco, F. B., et al. 2013, PASP, 125, 1031

\bibitem[Bryson et al.(2010)]{bryson10} Bryson, S.T., Tenenbaum, P., Jenkins, J.M., et al. 2010, ApJ, 713, 97

\bibitem[Cakirli et al.(2010)]{Cakirli} Cakirli, \"O, Ibanuglu, C., \& Dervisoglu, A. 2010, RMxAA, 46, 363 

\bibitem[Carter et al.(2011)]{carteretal11} Carter, J. A., Fabrycky, D. C., Ragozzine, D. et al., 2011, Science, 331, 562

\bibitem[Castelli \& Kurucz(2004)]{castellikurucz04} Castelli, F., \& Kurucz, R.~L., 2004, arXiv:astro-ph/0405087

\bibitem[Chambers et al.(2016)]{chambers16} Chambers, K.C., Magnier, E.A., Metcalfe, N., et al. 2016, arXiv:1612.05560

\bibitem[Christiansen et al.(2108)]{christiansen} Christiansen, J.L., Crossfield, I.J.M., Barensten, G., et al.~2018, AJ, 155, 57

\bibitem[Chubak et al.(2012)]{Chubak12} Chubak, C., Marcy, G., Fischer, D.A., Howard, A.W., Isaacson, H., Johnson, J.A., \& Wright, J.T. 2012, arXiv:1207.6212 

\bibitem[Claret \& Bloemen(2011)]{claretbloemen11} Claret, A., \& Bloemen, S., 2011, \aap, 529, A75

\bibitem[Collins et al.(2017)]{Collins} Collins, K.A., Kielkopf, J. F., Stassun, K. G., \& Hessman, F. V. 2017, AJ, 153, 77

\bibitem[Conroy et al.(2014)]{conroyetal14} Conroy, K.~E., Pr\v{s}a, A., Stassun, K.~G., Orosz, J.~A., Fabrycky, D.~C., \& Welsh, W.~F. 2014, \aj, 147, 45

\bibitem[Cutri et al.(2013)]{Cutri} Cutri, R.M., Wright, E.L., Conrow, T., et al.~2013, wise.rept, 1C.

\bibitem[David et al.(2018)]{david} David, T.J., Crossfield, I.J.M., Benneke, B., et al.~2018, AJ, 155, 222

\bibitem[Derekas et al.(2011)]{derekasetal11} Derekas, A., Kiss, L. L., Borkovits, T. et al., 2011, Science, 332, 216

\bibitem[Dittmann et al. (2017)]{Dittmann} Dittmann, J.A., Irwin, J.M., Charbonneau, D., et al. 2017, \apj, 836, 124

\bibitem[Eggleton \& Kiseleva-Eggleton(2001)]{eggletonkiseleva-eggleton01} Eggleton, P. P., \& Kiseleva-Eggleton, L., 2001, \apj, 562, 1012

\bibitem[Eggelton et al.(1998)]{eggletonetal98} Eggleton, P. P., Kiseleva, L., Hut, P., 1998, \apj, 499, 853

\bibitem[Endl \& Cochran (2016)]{endl_cochran16} Endl, M., \& Cochran, W.D. 2016, PASP, 128, 94502

\bibitem[Fabrycky \& Tremaine(2007)]{fabryckytremaine07} Fabrycky, D., \& Tremaine, S., 2007, \apj, 669, 1298


\bibitem[Fehlberg(1974)]{fehlberg74} Fehlberg, E., 1974, NASA Technical Report R-432

\bibitem[Flower(1996)]{flower96} Flower, P. J., 1996, \apj, 469, 355

\bibitem[Ford(2005)]{Ford} Ford, E.B. 2005, \aj, 129 1706

\bibitem[\protect\citeauthoryear{Gim\'enez \& Garcia-Pelayo}{1983}]{gimenezgarcia83}
 Gim\'enez, A., Garcia-Pelayo, J. M., 1983, \apss, 92, 203

\bibitem[Haas et al.(2012)]{haasetal12} Haas, M., Hackstein, M., Ramolla, M. et al., 2012, AN, 333, 706

\bibitem[Hajdu et al.(2017)]{hajduetal17} Hajdu, T., Borkovits, T., Forgács-Dajka, E. et al., 2017, \mnras, 471, 1230

\bibitem[Han et al.(2017)]{hanetal17} Han, E., Muirhead, Ph. S., Swift, J. J. et al., 2017, \aj, 154, 100 

\bibitem[He{\l}miniak et al.(2017)]{helminiaketal17} He{\l}miniak, K. G., Ukita, N., Kambe, E., Koz{\l}owski, S. K., Paw{\l}aszek, R., Maehara, H., Baranec, C., Konacki, M., 2017, \aap, 602, A30

\bibitem[Howard et al.(2010)]{Howard10} Howard, A. W., Johnson, J. A., Marcy, G. W., et al. 2010, ApJ, 721, 1467

\bibitem[Howell et al. (2014)]{howell14} Howell, S.B., Sobeck, C., Hass, M., et al. 2014, \pasp, 126, 398

\bibitem[Irwin(1959)]{irwin59} Irwin, J. B., 1959, \aj, 64, 149

\bibitem[Kipping et al.(2015)]{Kipping} Kipping, D. M., Schmitt, A. R., Huang, X., Torres, G., Nesvorn\'y, D., Buchhave, L. A., Hartman, J., \& Bakos, G. \'A. 2015, ApJ, 813, 14

\bibitem[\protect\citeauthoryear{Kiseleva et al.}{1998}]{kiselevaetal98}
 Kiseleva, L. G., Eggleton, P. P., Mikkola, S., 1998, \mnras, 300, 292
 
 \bibitem[Kolbl et al.(2015)]{Kolbl15} Kolbl, Rea, Marcy, G. W., Isaacson, H., \& Howard, A. W. 2015, AJ, 149, 18


\bibitem[Kraus et al.(2011)]{Kraus} Kraus, A.L., Tucker, R.A., Thompson, M.L., Craine, E.R., \& Hillenbrand, L.A. 2011, \apj, 728, 48


\bibitem[Lindegren et al.(2018)]{lindegren} Lindegren, L., Hernandez, J, Bombrun, A., et al. 2018, arXiv:1804.09366.

\bibitem[Loeb \& Gaudi(2003)]{loebgaudi03} Loeb, A., \& Gaudi, B.S. 2003, \apj, 588, 117

\bibitem[Lucy(1967)]{lucy67} Lucy, L. B., 1967, Zeitschrift f\H ur Astrophisik, 65, 89

\bibitem[Maoz et al.(2014)]{maozetal14} Maoz, D., Mannucci, F., \& Nelemans, G., 2014, ARA\&Ap, 52, 107

\bibitem[Marsh et al.(2014)]{marshetal14} Marsh, T.R., Armstrong, D.J., Carter, P.J., 2014, \mnras, 445, 309

\bibitem[Masuda et al.(2015)]{masudaetal15} Masuda, K., Uehara, Sh., \& Kawahara, H., 2015, \apj, 806, L37

\bibitem[Maxwell \& Kratter(2018)]{maxwellkratter18} Maxwell, M., Kratter, K. M., 2018, \apj, 854, 44


\bibitem[Naoz \& Fabrycky(2014)]{naozfabrycky14} Naoz, S., \& Fabrycky, D. C., 2014, \apj, 793, 137


\bibitem[Nemravov\'a et al.(2016)]{nemravovaetal16} Nemravov\'a, J. A. et al., 2016, \aap, 594, A55

\bibitem[Orosz(2015)]{orosz15} Orosz, J. A., 2015, ASPC, 496, 55

%
\bibitem[Perets \& Fabrycky(2009)]{peretsfabrycky09} Perets, H. B., \& Fabrycky, D. C., 2009, \apj, 697, 1048

\bibitem[Portegies Zwart et al.(2011)]{portegieszwartetal11} Portegies Zwart, S., van den Heuvel, E. P. J., van Leeuwen, J., Nelemans, G., 2011, \apj, 734, 55

\bibitem[Pr\v{s}a \& Zwitter(2005)]{Phoebe} Pr\v{s}a, A., \& Zwitter, T., 2005, \apj, 628, 426

\bibitem[Ransom et al.(2014)]{ransometal14} Ransom, S. M., Stairs, I. H., Archibald, A. M. et al., 2014, Nature, 505, 520 

\bibitem[Rappaport et al.(2017)]{Rappaport17} Rappaport, S., Vanderburg, A., Borkovits, T., et al. 2017, \mnras, 467, 2160

\bibitem[Skrutskie et al.(2006)]{Skrutskie} Skrutskie, M.F., Cutri, R.M., Stiening, R., et al. 2006, AJ, 131, 1163.

\bibitem[Shappee \& Thompson(2013)]{shappeethompson13} Shappee, B. J., \& Thompson, T. A., 2013, \apj, 766, 64

\bibitem[Shibahashi \& Kurtz(2012)]{TheThing} Shibahashi, H., \& Kurtz, D.W. 2012, \mnras, 422, 738

\bibitem[Slawson et al.(2011)]{slawsonetal11} Slawson, R. W., Pr\v sa A., Welsh et al., 2011, \aj, 142, 160

\bibitem[Stassun et al.(2018)]{Stassun18} Stassun, K.G., Oelkers, R.J., Pepper, J., et al. 2018, \aj, 156, 102


\bibitem[Tokovinin(2014)]{tokovinin14}  Tokovinin, A., 2014b, \aj, 147, 87

\bibitem[Tokovinin(2017)]{tokovinin17} Tokovinin, A., 2017, \apj, 844, 103

\bibitem[Tokovinin(2018)]{tokovinin18} Tokovinin, A., 2018, \aj, 155, 160

\bibitem[Tokovinin et al.(2006)]{tokovininetal06} Tokovinin, A., Thomas, S., Sterzik, M., Udry, S., 2006, \aap, 450, 681 

\bibitem[Torres(2010)]{torres10} Torres, G. 2010, \aj, 140, 1158

\bibitem[Tout et al.(1996)]{toutetal96} Tout, C.A., Pols, O.R., Eggleton, P.P., \& Han, Z. 1996, \mnras, 281, 257

\bibitem[Tull et al.(1995)]{tull95} Tull, R.G., MacQueen, P.J., Sneden, C., \& Lambert, D.L. 1995, PASP, 107, 251

\bibitem[Yu et al. (2018)]{yu} Yu, L., Crossfield, I.J.M., Schlieder, J.E., et al. 2018, arXiv:1803.0409

\bibitem[Vanderburg \& Johnson (2014)]{vanderburg2014} Vanderburg, A., and Johnson, J. 2014, \pasp, 126, 948

\bibitem[van Kerkwijk et al.(2010)]{vankerkwijketal10} van Kerkwijk, M.H., Rappaport, S., Breton, R., Justham, S., Podsiadlowski, Ph., \& Han Z. 2010, ApJ, 715, 51

\bibitem[Villanueva et al.(2018)]{Villanueva18}Villanueva, S., Jr., Gaudi, B. S., Pogge, R. W., Eastman, J. D., Stassun, K. G., Trueblood, M., \& Trueblood, P. 2018, PASP, 130, 5001

\bibitem[Yee et al.(2017)]{Yee17} Yee, S.W., Petigura, E.A., \& von Braun, K. 2017, ApJ, 836, 77

\bibitem[Zacharias et al.(2013)]{UCAC4} Zacharias, N., Finch, C.T., Girard, T.M., Henden, A., Bartlett, J.L., Monet, D.G., \& Zacharias, M.I. 2013, ApJS, 145, 44

\end{thebibliography}




\appendix

\section{Some details of the numerical integrator and the initialization of the orbital elements}
\label{app:numint}

The numerical 3-body integrator which is built into our code was originally developed for studying tidal effects on the dynamics of hierarchical triple stellar systems. Hence, it integrates not only the equations of orbital motion of the three bodies, including tidal and rotational effects with arbitrarily directed stellar spin axes, but also, simultaneously, the Eulerian equations of rotation of the stellar components, regarding the stars as deformable bodies.\footnote{We stress that general relativistic effects, including relativistic precession, have not yet been included into the equations of motions. Note, however, that in our system the relativistic precession rates are $(\dot\omega_1)_\mathrm{GR}\approx19\farcs3$\,yr$^{-1}$ and $(\dot\omega_2)_\mathrm{GR}\approx0\farcs17$\,yr$^{-1}$, which can safely be ignored relative to the dynamical apsidal motion rate of $(\dot\omega_1)_\mathrm{dyn}\approx15\fdg3$\,yr$^{-1}$.} The dependent variables of the integrator are the six Cartesian coordinates and velocity components of the first two Jacobian vectors (i.e., $\vec{\rho}_1$ which connects the center of mass of the two components of the inner binary, directed from the primary to the secondary, and $\vec{\rho}_2$ which connects the center of mass of the inner binary with the outer, third component), and also the Eulerian angles and their first derivatives, describing the stellar rotations for each star individually. The details of the integrator were given in \citet{borkovitsetal04}, and we consider here only some relevant notes on the conversion between the orbital elements and the Cartesian inputs and outputs of our integrator in the case of such a compact, dynamically active triple system.

\begin{figure*}
\begin{center}
\includegraphics[width=0.32 \textwidth]{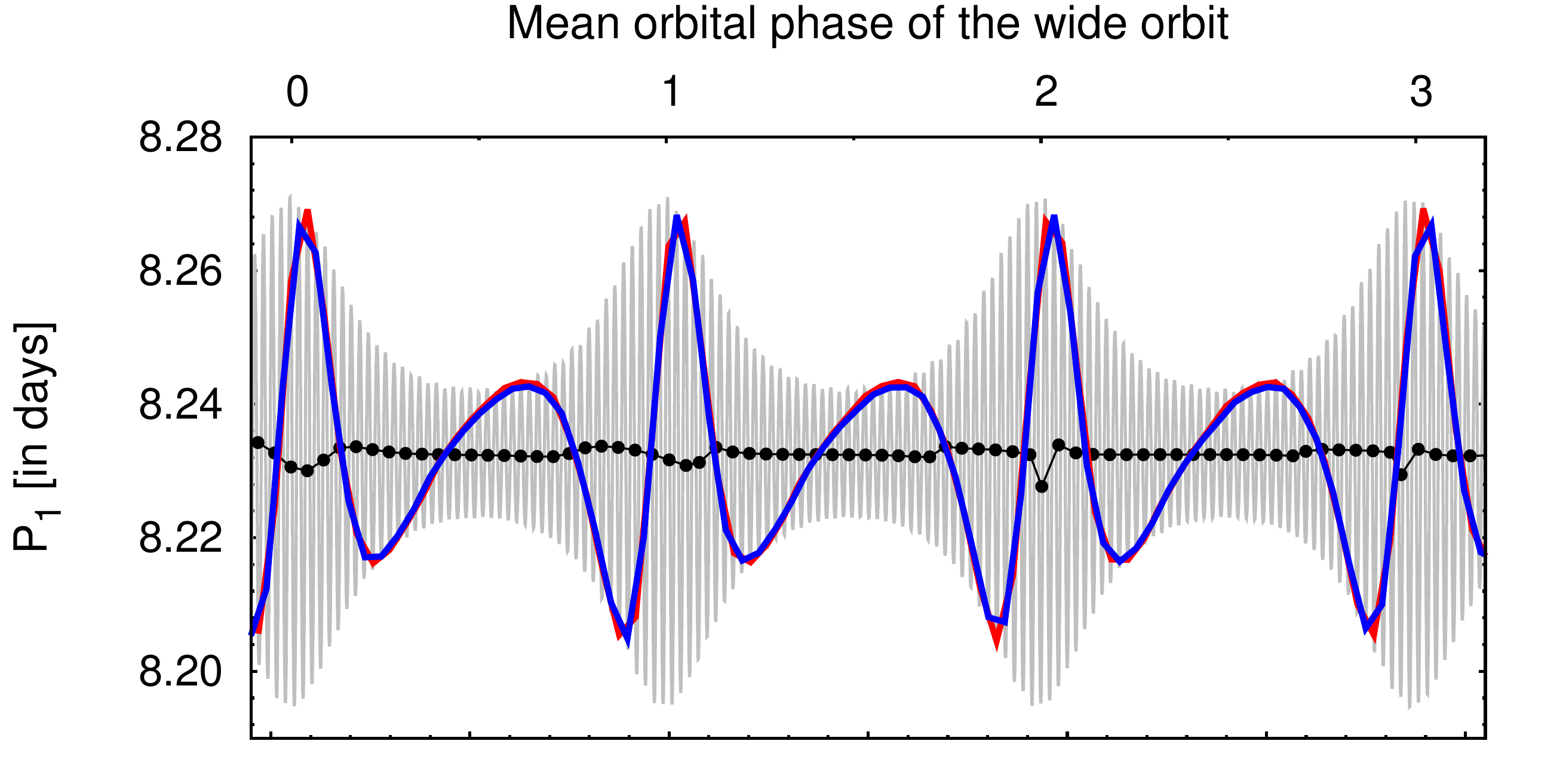}\includegraphics[width=0.32 \textwidth]{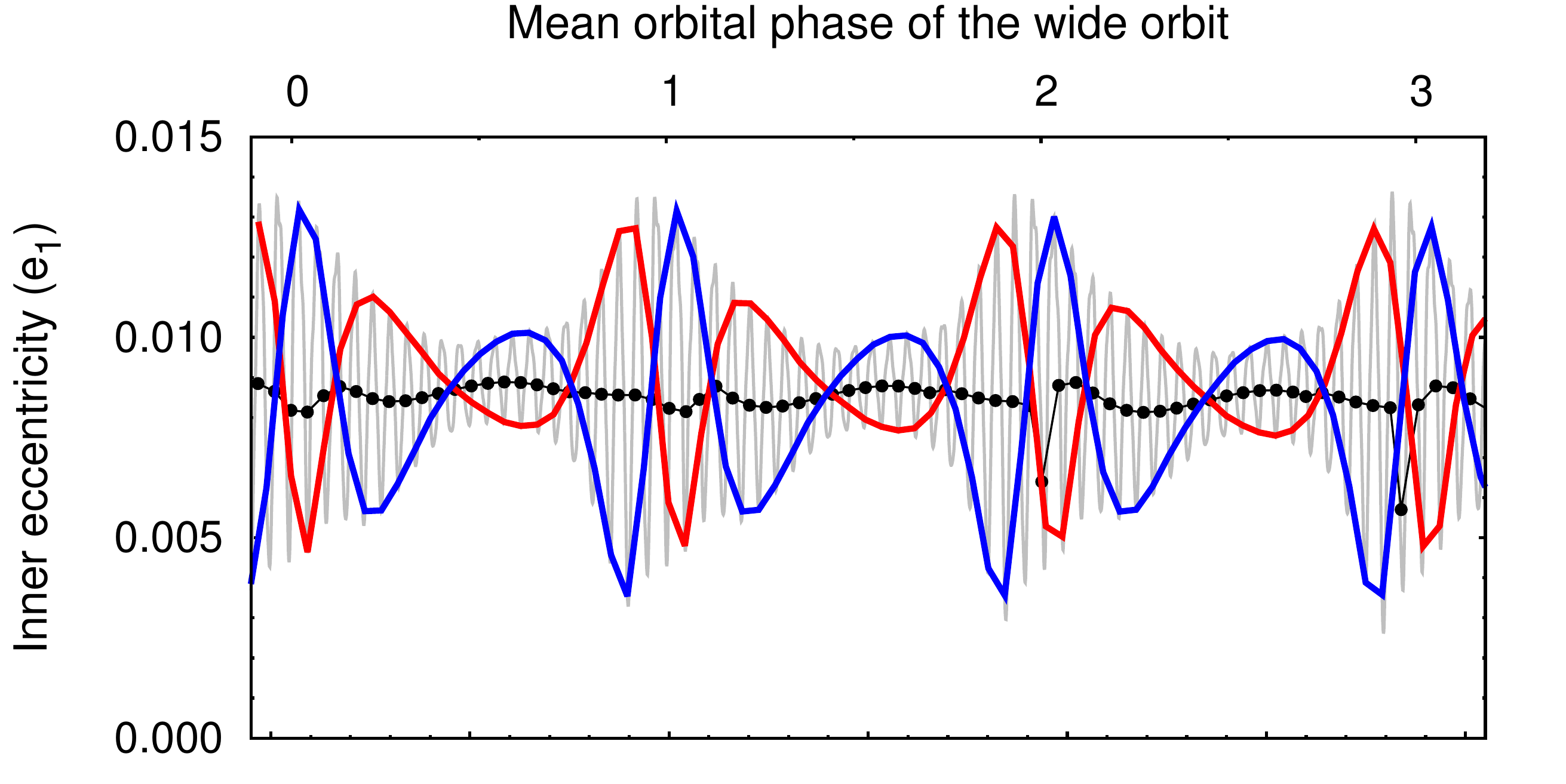}\includegraphics[width=0.32 \textwidth]{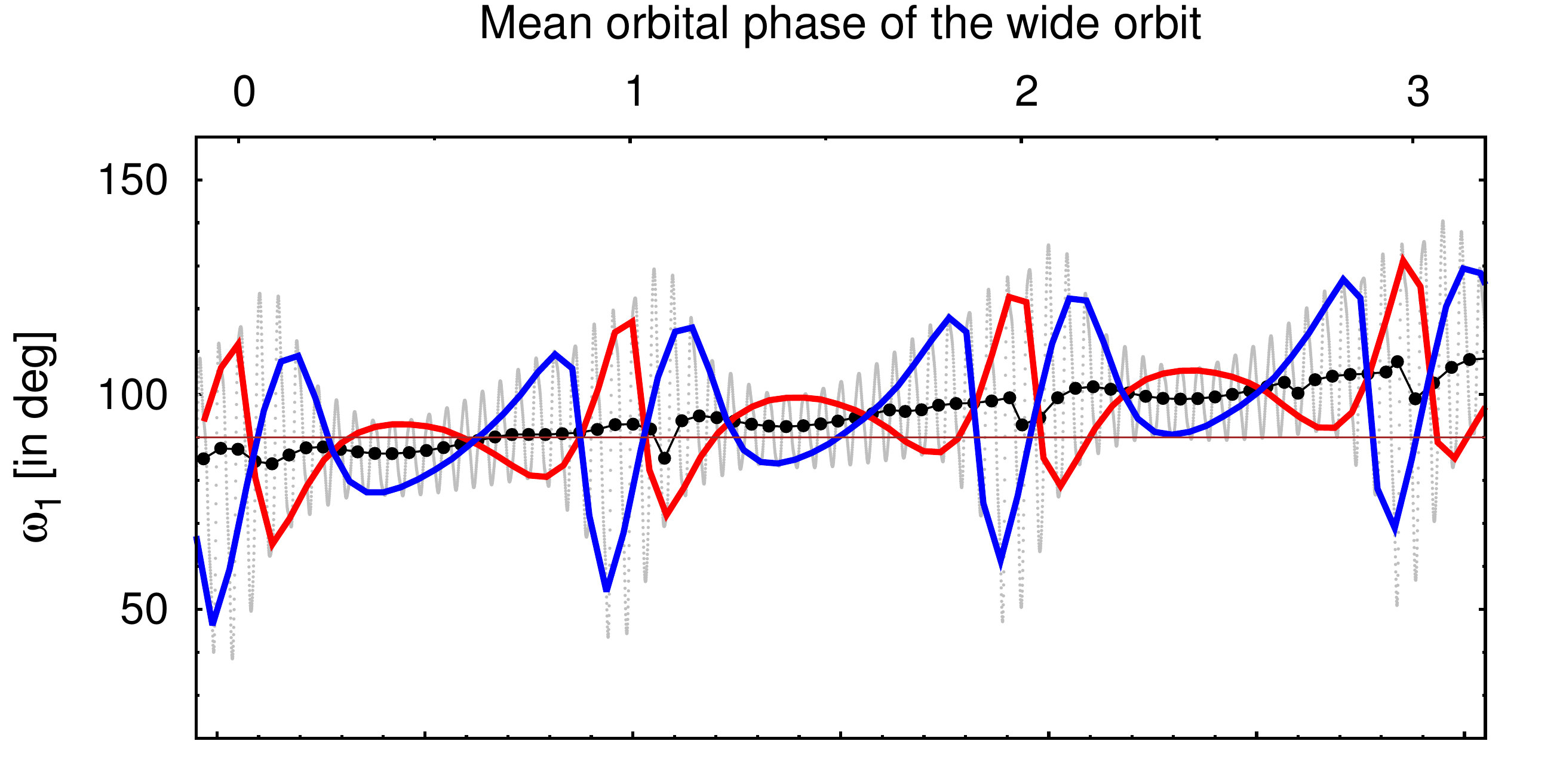}
\includegraphics[width=0.32 \textwidth]{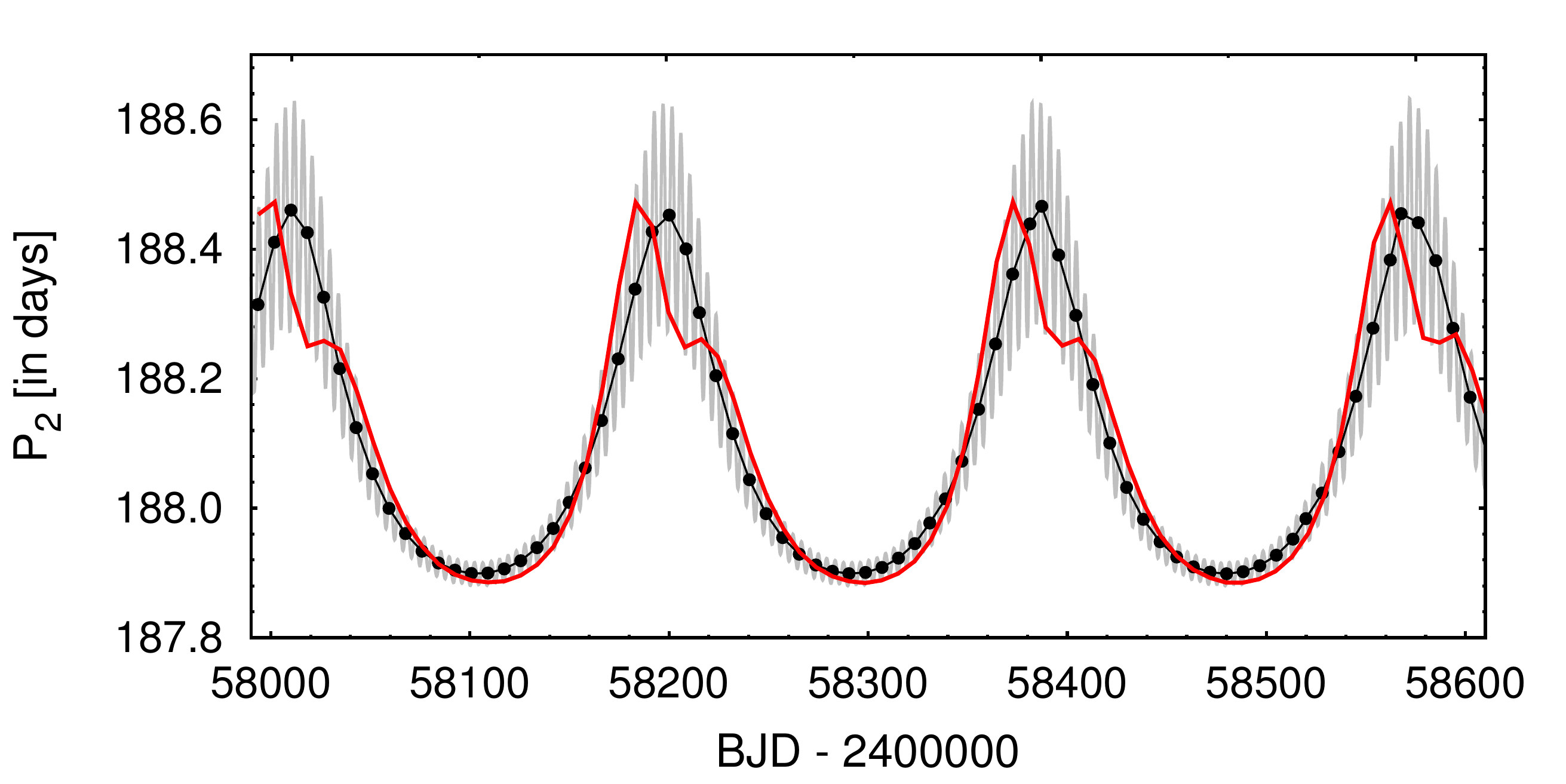}\includegraphics[width=0.32 \textwidth]{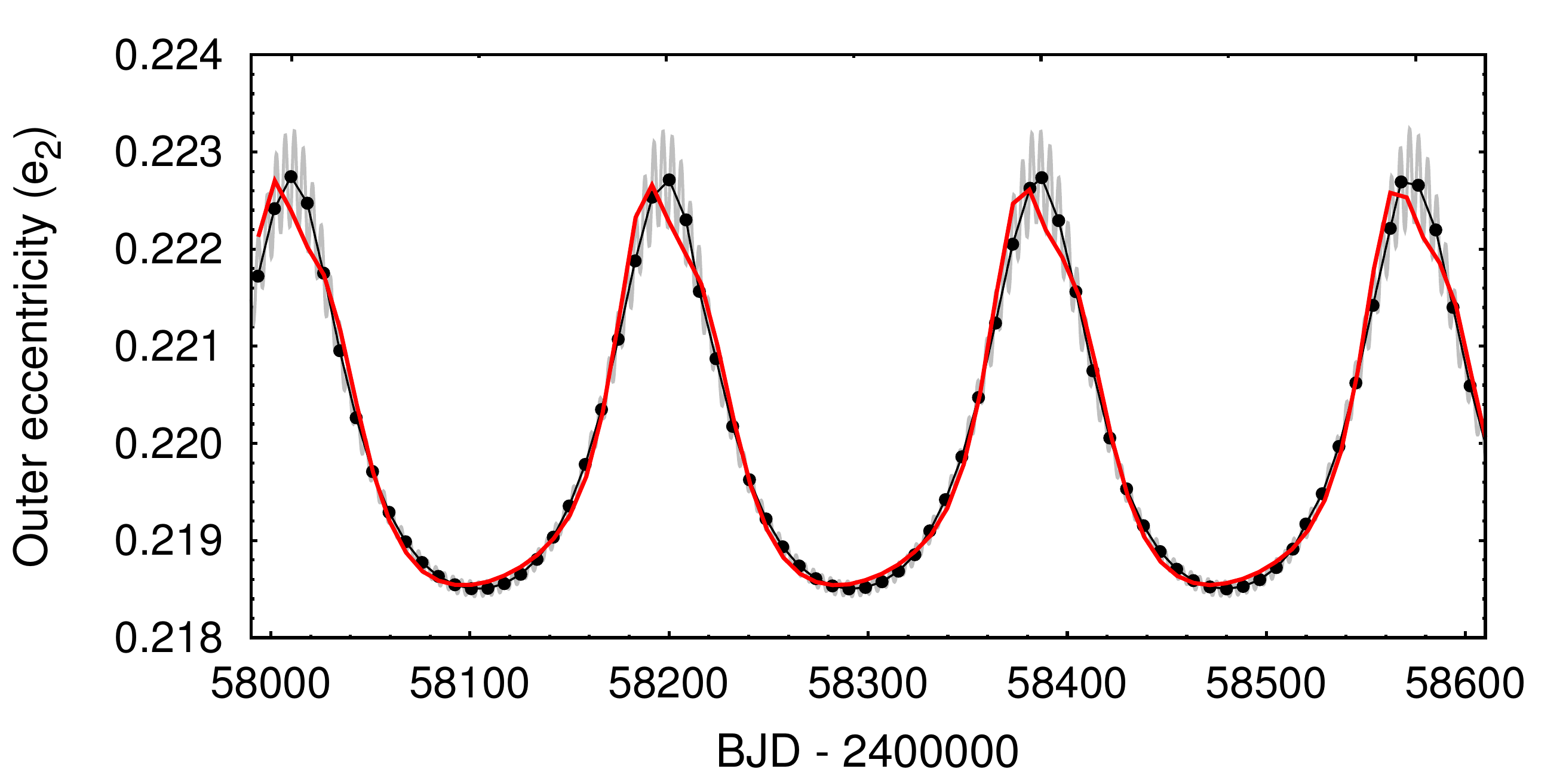}\includegraphics[width=0.32 \textwidth]{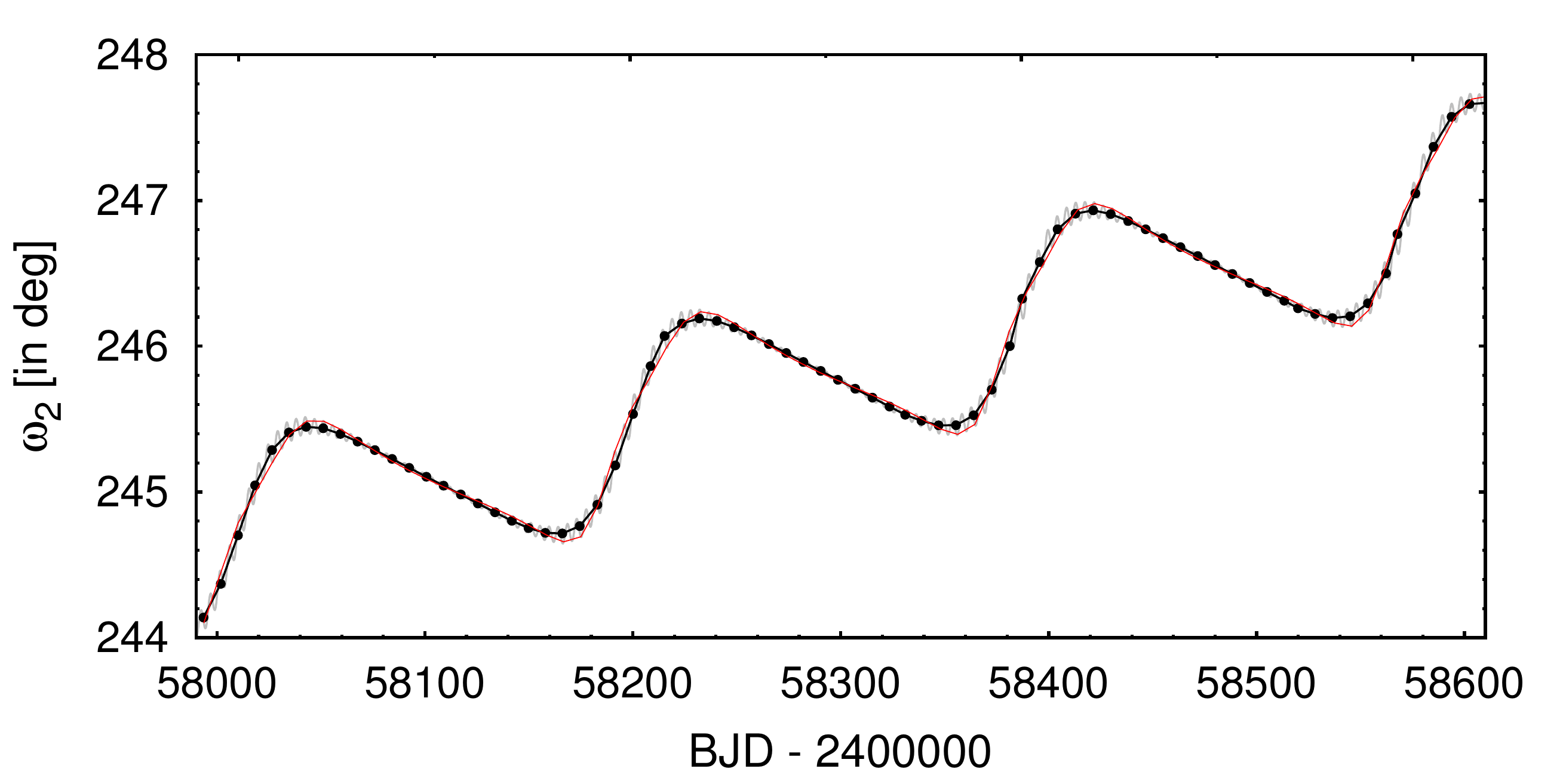}
\caption{Variations of the instantaneous (osculating) anomalistic periods, eccentricities and (observable) arguments of periastron of the inner and outer binaries (grey curves in all panels), obtained via numerical integration, together with the one-inner-orbit averages (black).  Furthermore, the red and blue lines connect the instantaneous orbital elements sampled at those integration points that are closest to the primary and secondary mid-eclipse points of the inner binary, respectively. (Note, for the outer orbit the red and blue lines coincide almost perfectly, therefore, we plot only the red one.) The brown horizontal line at $\omega_1=90\degr$ (in the upper right panel) helps to identify the domain where the phase displacement of the inner secondary eclipse relative to photometric phase $0\fp5$ reverses its sign. (See text for details.)}
\label{fig:orbelements_numint} 
\end{center}
\end{figure*}

As is well known, in the case of either third-body or tidal perturbations, the orbital motion of a binary component will no longer be pure Keplerian and, therefore, the usual orbital elements will not retain their exact meaning.  There are, however different treatments to preserve and generalize the meanings of orbital elements for perturbed systems.  In our photodynamical code we do this in the most usual and intuitive way, i.e. we use the so-called `osculating' orbital elements. This means that the orbital elements (used only as input parameters) are converted to Cartesian coordinates and velocities at the given epoch in the same way as for the unperturbed case. One should keep in mind, however, that osculating orbital elements at later instants of time will differ from the initial one, as is nicely illustrated in Fig.\,\ref{fig:orbelements_numint}. This requires some extra care when one intends to compare the photodynamical results with some observational results obtained at significantly different epochs.

Furthermore, Fig.\,\ref{fig:orbelements_numint} illustrates some other issues, both theoretical and practical ones. For example, as one can see (in the upper row), for very small binary eccentricities, the argument of periastron is very weakly determined.  In other words, the orientation of the major axis of the orbital ellipse might be subject to large variations within very short intervals. This fact reveals a principal weakness in using the osculating orbital elements. In order to illustrate this, let us consider a point mass orbiting a body that is a permanently deformed into an oblate spheroid (e.g., rotationally deformed) on an exactly circular orbit.  In this case, as long as the extra (tidal) force remains radial and constant in time, the motion remains strictly circular.  On the other hand, calculating the osculating orbital elements of the point mass, one would infer a small, but definitely non-zero, eccentricity and a continuously rotating argument of periastron due to the fact that the orbiting point mass is located permanently at the periastron point of the osculating ellipse. (This question was discussed, and some suggestions for resolving the problem were given, e.g., in \citealt{kiselevaetal98} and \citealt{borkovitsetal02}.) Fortunately, as we found, the use of the alternative orbital elements $e\cos\omega$ and $e\sin\omega$ (or, more or less equivalently, the displacement of the secondary eclipse from photometric phase $0\fp5$) during the initialization of the numerical integrator avoids this problem, therefore, it has no practical consequence.

By contrast, the large fluctuations in the osculating instantaneous anomalistic period of the inner pair (top left panel) have very unpleasant consequences for the speed of convergence of the MCMC parameter search; this is especially true when the ETV curves are included into the fitting process. The reason is as follows. It is well known that the ETV is extraordinarily sensitive to the mean sidereal period since the amplitude of each ETV point is proportional to the product of the error in that period and the cycle number, growing linearly in time. Here the following problem arises. The relevant input parameter of the photodynamics code is the instantaneous anomalistic period; however, the ETV is governed chiefly by the mean sidereal period. The instantaneous anomalistic period depends strongly on the orbital elements of both orbits, and also on the accurate locations of the stars along their orbits (i.e. the orbital phases), not to mention their masses. Therefore, small changes in any of these parameters at each trial step in the fitting process lead to a significant mis-tuning of the mean sidereal period, thereby resulting in large $\chi_\mathrm{ETV}^2$ values.  This can slow significantly the convergence of the MCMC chains despite the fact that the fits were seemingly good anyway. 

The ideal solution to the mismatch between the two different types of periods would be to try to convert the mean sidereal period to the instantaneous anomalistic period at the epoch when the numerical integrator is initialized and, thereby, the former would be used as an input parameter. To find such an analytical relation would require, however, unrealistically large and time-consuming efforts, and thus we applied another simple, heuristic solution, as follows. During each trial step, our code first calculates the numerical ETV points (see Appendix\,\ref{app:ETV} below) and then adjusts this numerical ETV to the observed one with a two-parameter linear least-squares fit using the difference in the epoch ($T_0$) and mean sidereal period ($P_\mathrm {s}$) as free parameters. Then these differences are considered as approximate corrections to the initial instantaneous anomalistic orbital period and the initial phase-term of the inner orbit, and they are therefore simply added to the previous input values. Then, the same process is repeated once again (we found that two iterations   satisfactorily constrain these parameters), and the resultant initial parameters are used for the subsequent trial step of the given chain.  This includes the calculation not only of the model lightcurves and RV curve, but also the re-calculation of the model ETVs.

Finally, for the sake of completeness, we discuss another issue regarding the initialization of certain parameters at each trial step. This is the question of the synchronous rotation of any of the stars. Regarding the orbital motion of a binary system, it is well-known that in the presence of any third-body perturbations, the orbit cannot remain permanently exactly circular, and therefore not exactly synchronous. Furthermore, unless the perturber is exactly located in the binary's orbital plane, there should also occur a precession of that plane. As a consequence, one also can expect that the equatorial plane of the binary stars and the outer orbital plane will no longer be continuously aligned.  Therefore, due to an inclined perturber, some stellar spin precession should also occur (for further discussion see e.\,g. \citealp{eggletonetal98,kiselevaetal98,borkovitsetal04}).

Taking into account these considerations in our numerical integrator (and, therefore, in the complete photodynamical treatment) we apply the following iterative process during the initialization of the spin parameters of a star flagged as a `quasi-synchronous rotator':
\begin{itemize}
\item[(i)]{First the code takes the stellar rotational equatorial plane to be identical with the orbital plane, setting the Eulerian angles of stellar rotation equal to the corresponding orbital elements\footnote{Orbital elements $i$, $\Omega$ are also practically Eulerian angles, describing the orientation of the orbital plane relative to the base of the coordinate system, i.e. the tangential plane of the sky.} as $\theta=i$, $\phi=\Omega$, and setting the uni-axial spin angular velocity equal to the orbital angular velocity calculated at periastron passage, i.e. $\omega_{z'}=w_\mathrm{per}$.}
\item[(ii)]{Using these initial values, the code calculates the instantaneous Cartesian force components.}
\item[(iii)]{Using the known perturbing force, the code calculates the instantaneous time derivatives of the angular orbital elements, and sets $\dot\theta=\dot\iota$ and $\dot\phi=\dot\Omega$. Similarly, the perturbed orbital angular velocity (at periastron passage) is calculated as well and, the uni-axial spin angular velocity (and, similarly, the time derivative of the third Eulerian angle, $\dot\psi$) are set accordingly.}
\item[(iv)]{Finally, the same process from item (ii) is repeated twice again.}
\end{itemize}

\section{Numerical ETV curve generation}
\label{app:ETV}

Both photometric fluxes and radial velocities at any given time are direct outputs of the lightcurve and RV curve emulator code for any sets of the initial model parameters, and therefore, can be readily compared to their observational counterparts.  However, this is not so for the model times of eclipse minima.  In order to obtain the necessary mid-eclipse times and, therefore, to obtain the model ETV points from the numerical integration of the equations of motion of the triple system, we applied the following quick and approximate, but sufficiently accurate method. 

The numerical integrator calculates the Jacobian coordinates and velocities of the three stars at any given observed mid-eclipse time ($t_\mathrm{obs}$).  Then, these coordinates and velocities were converted back to instantaneous osculating orbital elements, as well as the corresponding instantaneous eccentric anomaly, $\mathcal{E}_1(\mathrm{t_{obs}})$, which we write more simply as $\mathcal{E}_{\rm 1, obs}$, where the subscript ``{\footnotesize 1}'' refers to the inner binary, as opposed to the outer orbit.  By the use of the same osculating, instantaneous orbital elements, the theoretical true anomaly at the moment of the model mid-eclipse was also calculated as 
\begin{equation}
f_\mathrm{1,calc}=\mp\frac{\pi}{2}-\omega_1\pm\frac{e_1\cos\omega_1\cos^2i_1}{\sin^2i_1\mp e_1\sin\omega_1},
\label{Eq:fmin}
\end{equation}
where the last term on the right side is a first-order approximation of the exact expression given by \citet{gimenezgarcia83} describing the very weak inclination-angle dependence of the mid-eclipses for eccentric orbits. The upper signs refer to primary eclipses, while the lower signs are for the secondary ones. (Note that we define $\omega$ as the argument of periastron of the {\em secondary} component relative to the primary.)  Then, we convert this true anomaly at the time of a model eclipse to the corresponding eccentric anomaly via the familiar relation
\begin{equation}
\mathcal{E}_\mathrm{1,calc}=2\arctan\left(\sqrt{\frac{1-e_1}{1+e_1}}\tan\frac{f_\mathrm{1,calc}}{2}\right),
\end{equation}
From $\mathcal{E}_\mathrm{1,calc}$ and $\mathcal{E}_{\rm 1, obs}$ the model mid-eclipse times, relative to the observed ones, can be calculated easily as
\begin{eqnarray}
t_\mathrm{model}\!&\!=\!&t_\mathrm{obs}-\frac{m_\mathrm{A}}{m_\mathrm{AB}}\frac{Z_2}{c}\nonumber \\
&\!+ &\!\frac{P_1}{2\pi}\left[\mathcal{E}_\mathrm{1,calc}-\mathcal{E}_\mathrm{1,obs}-e_1\left(\sin\mathcal{E}_\mathrm{1,calc}-\sin\mathcal{E}_\mathrm{1,obs}\right)\right]. \nonumber \\
\label{Eq:t_model}
\end{eqnarray}
Here, the last term in the first row of the right hand side stands for the light-travel time contribution, where $Z_2$ denotes the $Z$ coordinate of the second Jacobian vector, i.e., the radius vector directed from the centre of mass of the inner pair to the outer component.

We tested the accuracy of this fast, but approximate, calculation of the model eclipse times by comparing these values with those calculated from the model lightcurve by the same method as the observed eclipse times were calculated. We found good correspondance between the two types of simulated ETV data, down to a scale of $\sim$$10^{-4}$\,days.


\bsp	
\label{lastpage}
\end{document}